\documentclass[aps,prl,reprint,superscriptaddress]{revtex4-2}
\usepackage{graphicx}
\usepackage{mathpazo}
\usepackage{dcolumn}
\usepackage{xcolor}
\usepackage{bm}
\usepackage{relsize}
\usepackage{tcolorbox}
\usepackage{braket}
\usepackage{lipsum}
\usepackage{amsmath, amsthm, amsfonts, amssymb, bbm}
\usepackage{mathtools}
\usepackage{pifont}
\usepackage[colorlinks,linkcolor=blue,urlcolor=blue, citecolor=blue]{hyperref}

\newcommand{\ketbra}[2]{|#1\rangle\! \langle #2|}

\theoremstyle{definition}

\begin{document}

\preprint{APS/123-QED}

\title{A Temperature Change can Solve the Deutsch - Jozsa Problem :\\  \smaller{}\texttt{An Exploration of Thermodynamic Query Complexity}}

\author{Jake Xuereb}
\affiliation{Vienna Center for Quantum Science and Technology, Atominstitut, TU Wien, 1020 Vienna, Austria}
\email{jake.xuereb@tuwien.ac.at}

\date{October 15, 2025}

\begin{abstract}
We demonstrate how a single heat exchange between a probe thermal qubit and multi-qubit thermal machine encoding a Boolean function, can determine whether the function is balanced or constant, thus providing a novel thermodynamic solution to the Deutsch-Jozsa problem. We introduce a thermodynamic model of quantum query complexity, showing how qubit thermal machines can act as oracles, queried via heat exchange with a probe. While the Deutsch-Jozsa problem requires an exponential encoding in the number of oracle bits, we also explore a restricted Bernstein-Vazirani problem, which admits a linear thermal oracle and a single thermal query solution. We establish bounds on the number of samples needed to determine the probe temperature encoding the solution for the Deutsch-Jozsa problem, showing that it remains constant with problem size. Additionally, we propose a proof-of-principle experimental implementation to solve the 3-bit Bernstein-Vazirani problem via thermal kickback. This work bridges thermodynamics and complexity theory, suggesting that quantum thermodynamics could provide an unconventional route to computing beyond classical computation.
\end{abstract}

\maketitle

\textbf{Introduction} Quantum decision problems and the development of the quantum query complexity model~\cite{deutsch1992rapid,berthiaume_brassard_92,berthiaume_brassard_94,bernstein_vazirani,Watrous2009} served as seminal points in quantum computation, exemplifying clearly the difference between the computational power of quantum and classical physics. These models, which eventually led to the development of quantum algorithms, made a number of assumptions in what constitutes a \textit{quantum query} i) qubits are initialised and used in pure states ii) unitaries create coherence as a resource for querying. 

In quantum thermodynamics~\cite{anders_review,Goold_2016,binder2018thermodynamics} these assumptions are often impossible to satisfy--pure states require infinite energy to obtain with certainty~\cite{taranto_23}, or are unnecessary--a system may be cooled efficiently without the generation of coherences~\cite{hovhanisyan_13}. Indeed limited models of quantum computation like DQC1 circuits~\cite{DQC1,ambainis2000computing,laflamme2002introduction,non_clean} are able to solve classically hard problems~\cite{integrability,poulin_2,shepherd2006computation,shor2008estimating} without satisfying some of these assumptions. Recently, even computational models that make use of solely stochastic thermodynamics have been shown to provide advantage over conventional classical models of computation in some tasks~\cite{aifer2024thermodynamic,bartosik2024thermodynamicalgorithmsquadraticprogramming}.

\begin{figure}[h]
    \centering
    \includegraphics[width=\linewidth]{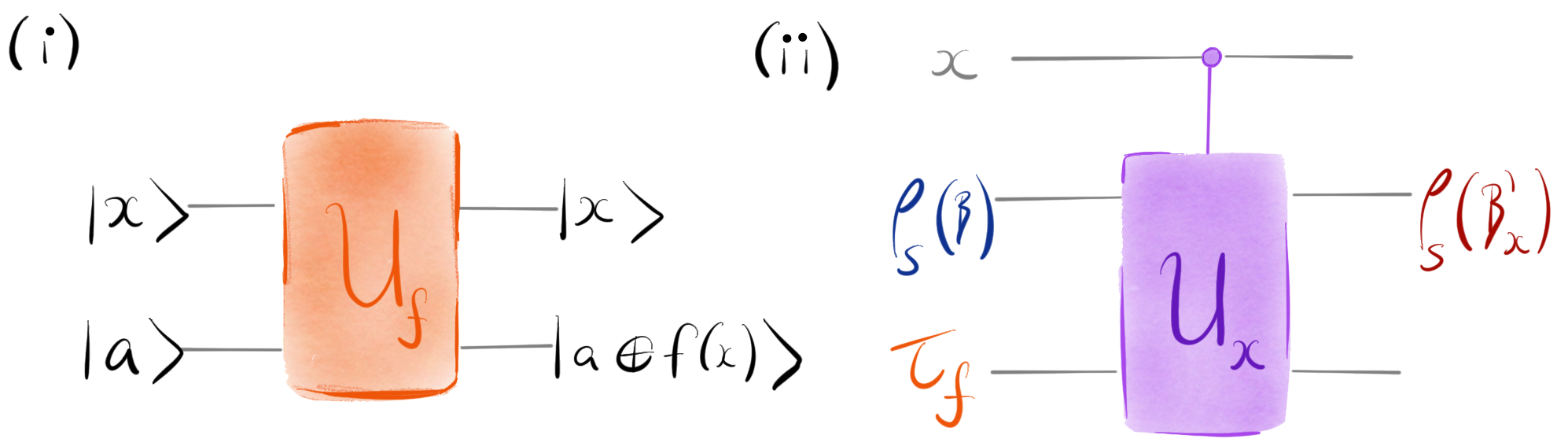}
    \caption{(i) A unitary binary oracle takes as input an $n$-qubit string $\ket{x} \,:\, x\in \{0,1\}^{\times n}$ and a single ancilla qubit $\ket{a} \, : \, a \in \{0,1\}$ imprinting the output of the Boolean function encoded by the oracle into $U_f$ onto the ancilla as $\ket{a \oplus f(x)}$. (ii) A thermal machine oracle $\tau_f$ is a collection of qubits whose energetic structure encodes $f$. An agent with access to a \textit{probe} $\rho_{S}(\beta)$ can exchange heat with $\tau_f$ via $U_x$ depending on an input $n$-bitstring $x$. The probe state changes to $\rho_S(\beta'_x)$ where the temperature of $S$ encodes the output $f(x)$.}
    \label{fig:oracles}
\end{figure}

This provokes us to wonder \textit{can a classically hard quantum decision problem be solved in a quantum thermodynamic scenario}? In this work we answer this question in the affirmative. We show that an agent with access to a thermal probe qubit can learn properties of a Boolean function, encoded by an oracle into the energetic gaps of a set of qubits forming a thermal machine~\cite{skrzypczyk_10,ronnie_12,ralph_swap,Mitchison_review}-- via a single heat exchange. We begin by examining how oracles for Boolean functions can be embedded into thermal machines and queried via different inputs which result in different heat exchanges with the probe. We investigate three different queries possible via heat exchange in our model (Fig.\ref{fig:oracles}) for the Deutsch-Jozsa~\cite{deutsch1992rapid} problem. In particular we present the \textit{thermal kickback} subroutine which imprints global properties of a function in the temperature of a probe qubit after a single heat exchange with the oracle's machine. We also show how the thermal kickback can detect the Hamming weight of an $n$-bit secret string. Following this we explore the thermodynamic properties the probe must have to be sensitive to a thermal kickback, giving an insight into the thermodynamic cost of querying quantum oracles. Importantly, we examine the number of copies of the probe that are required to determine its temperature and show that for the Deutsch-Jozsa problem the no. of samples required is constant with $n$ the size of the problem.  Finally, we discuss the implications of our work and present a schematic of an experimental implementation of our model for the $3$-bit Bernstein-Vazirani problem in the End Matter.

\textbf{From Unitary Oracles to Thermal Machine Oracles}    
In the traditional quantum query complexity setting~\cite{Watrous2009} an oracle with access to a Boolean function $f~:~\!\!\{0,1\}^{\times n}~\!\!\longrightarrow\!\!~\{0,1\}$ constructs a blackbox unitary Fig.\ref{fig:oracles}~(i) $U_f$ which maps an $n+1$ qubit input string $\ket{x \, a}$ to $\ket{x \, a\oplus f(x)}$ where the ancilla qubit's state now encodes the output of the query. Let us instead allow the oracle the ability to thermalise qubits constructing a thermal machine whose energetic structure encodes information about the output of the function. For each input string $x$ the oracle calls the function and prepares the state 
\begin{gather}
    \tau_x = \frac{1}{\mathcal{Z}_x}\left(\ketbra{0}{0} + e^{-\beta_M (f(x)E_1 + (f(x)\oplus 1)  E_2)} \ketbra{1}{1}\right),
\end{gather}
using the notation $E(x) = f(x)E_1 + (f(x)\oplus 1)E_2$ we have that the qubit Hamiltonian $H_x = 0 \ketbra{0}{0} + E(x)\ketbra{1}{1}$, the partition function $\mathcal{Z}_x = (1 + e^{-\beta_M E(x)})$ and the ground state energy of the qubit is set to 0. A detailed protocol for the oracle to prepare the thermal machine using conditional thermalisations is given in the supplemental material. Using a machine gap vector $\Gamma = (E(0^{N}), E(0^{N-1}1),E(0^{N-2}10), \dots, E(1^N))$ with entries corresponding to the energetic gap of the $x$th oracle qubit $E(x)$, the state of the \textit{thermal machine oracle} is then given by 
\begin{gather}
    \tau_f = \bigotimes_{x \in \{0,1\}^{\times n}} \tau_x = \sum_{i_M \in \{0,1\}^{\times n}} \frac{e^{-\beta_M i_M \cdot \Gamma}}{\mathcal{Z}_f}\ketbra{i_M}{i_M}, \label{eq:thermal_oracle}
\end{gather}
where $i_M\cdot \Gamma$ is the inner product between energy level $i_M$ and the gap vector $\Gamma$ giving the energy value of $i_M$. This construction allows for a thermodynamic query model Fig.\ref{fig:oracles}~(ii) where conditioned on an input bit string $x$ a heat exchange $U_x$ between a \textit{probe qubit} $\rho_S(\beta)$ initially at temp. $\beta = 1/T$ and the thermal machine oracle $\tau_f$ is carried out, altering the temperature of the probe to $\rho_S(\beta'_x)$ which encodes the output $f(x)$. A \textit{query} within this model is then a heat exchange between $\rho_S(\beta)$ and $\tau_f$ carried out by $U_x$. The output of this query is encoded in the temperature of a thermal state and so is \textit{probabilistic}, unlike the unitary query model where an eigenket is assumed as output, allowing for \textit{deterministic} output. This adds an additional complexity to the setting-- the \textit{sample complexity} i.e. no. of samples required to obtain an answer with a given error. Finally, note that $\tau_f$ has $2^n$ qubits in the size of the $n$-bit Boolean function. This is a simple oracle encoding, a classical truth table that can be queried quantum thermodynamically as a first toy model. 
Better oracle encodings likely exist e.g. the oracle could have set $H_M = \sum_{x \in \{0,1\}^{\times n}} f(x) \ketbra{x}{x}$ leading to an \textit{unphysical} linear encoding, but we leave investigating oracle encodings with low complexity for future work, focusing on this initial case. Despite its simplicity the presented encoding is still linear for some problems e.g.  Hamming weight detection we will present.

\begin{figure}
    \centering
    \includegraphics[width=\linewidth]{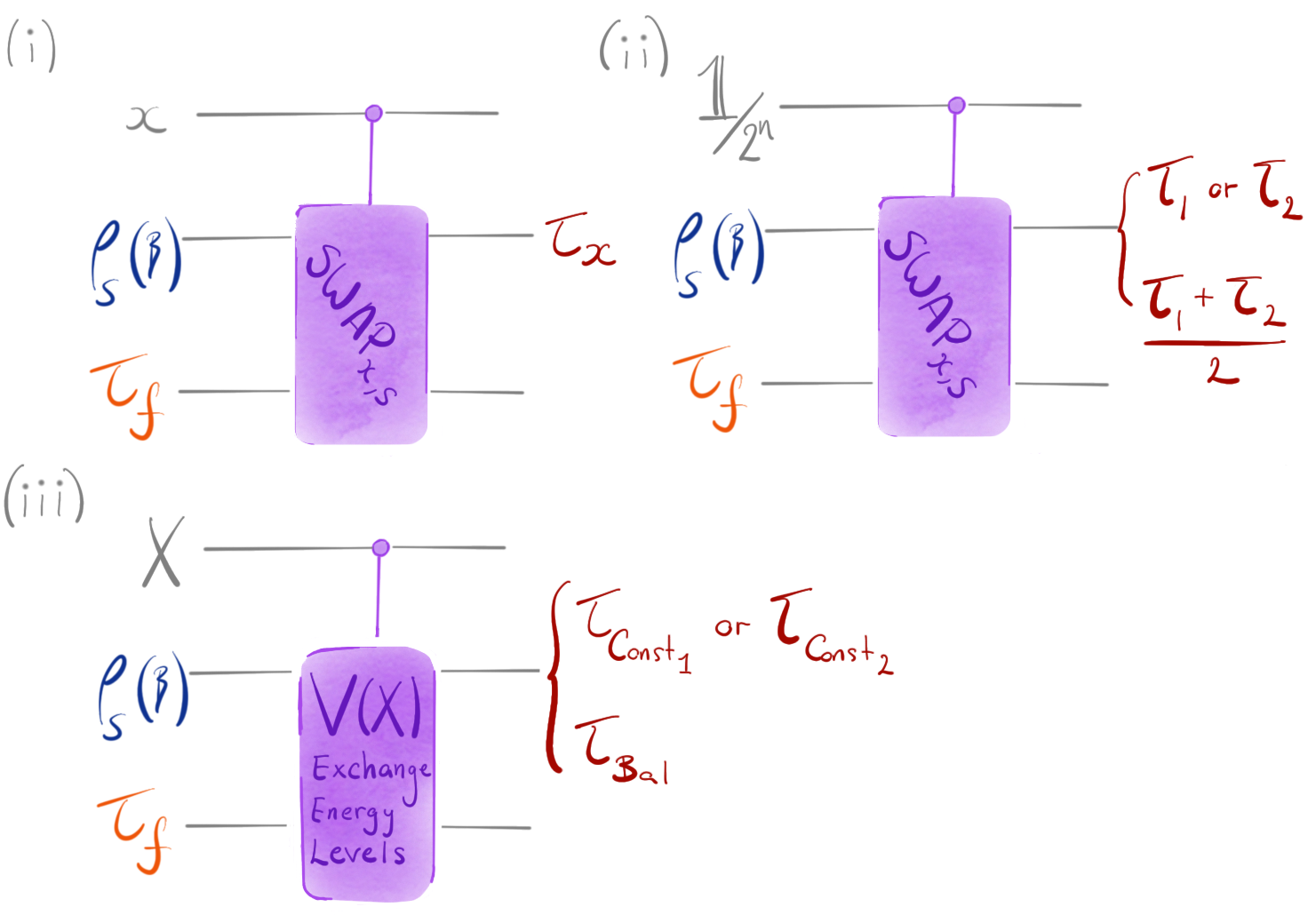}
    \caption{Three circuit diagrams depicting different heat exchanges across the probe qubit and thermal machine oracle which result in three different queries which are examined below for the Deutsch-Jozsa problem.}
    \label{fig:query_model}
\end{figure}

\begin{figure*}[t]
    \centering
    \includegraphics[width=\linewidth]{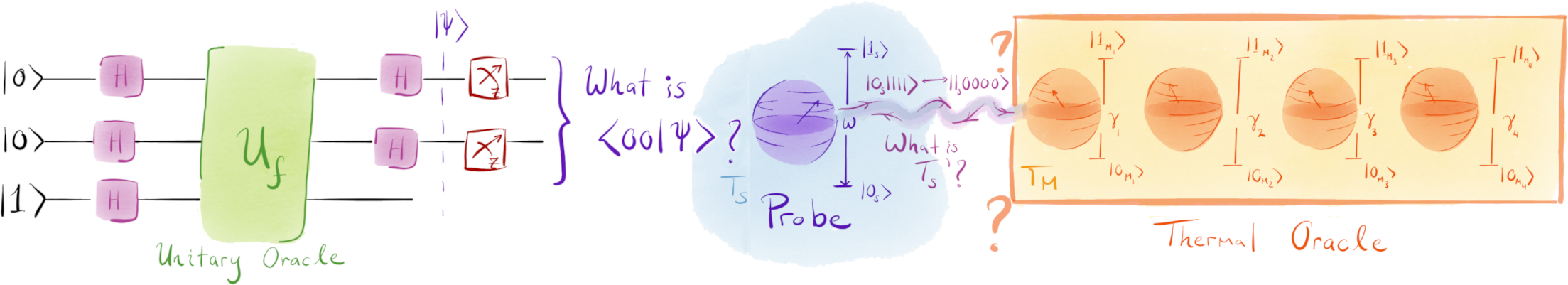}
    \caption{A visualisation of an illustrative example. At the left, we see the standard 3 qubit Deutsch-Jozsa circuit where an oracle of the 2-bit function $f(x)$ is implemented as a unitary and the decision problem is solved by evalutating $\braket{00|\psi}$. To the right, we see a 4 qubit thermal machine at temp. $T_M$ with gaps $\Gamma = (\gamma_1, \gamma_2, \gamma_3, \gamma_4)$ whose energy level structure is used by an oracle to encode $f(x)$, global properties of $f(x)$ can be determined in a single heat exchange with a probe at temp. $T_S$ with energetic gap $\omega$.}
    \label{fig:banner}
\end{figure*}

\textbf{The Deutsch - Jozsa Problem} To understand the core ideas behind the proposed thermodynamic query complexity model we begin by examining the Deutsch-Jozsa (DJ) problem and solve it via a thermodynamic kickback rather than a phase kickback. The $n$-bit Deutsch-Jozsa problem asks one to identify whether a Boolean function $f(x)~:~\{0,1\}^{\times n}~\rightarrow  ~\{0,1\}$ is constant, meaning for every $x \in \{0,1\}^{\times n}$ the output is always 0 or always 1, or balanced meaning the output is 0 for half the domain and 1 for the rest. To solve this problem, Deutsch and Jozsa~\cite{deutsch1992rapid} encoded the Boolean function $f(x)$ into a unitary oracle as in Fig~\ref{fig:oracles} (i). The problem is then solved using the circuit given in Fig~\ref{fig:banner} for the 2-bit DJ problem where two register qubits are put in a superposition over all possible strings and an ancilla qubit is made to interfere with this equal superposition of inputs while querying $U_f$. This imprints global information about $f(x)$ in terms of a \textit{phase} on the ancilla qubit, often referred to as a \textit{phase kickback}.

Consider instead a thermal oracle encoding $\tau_f$ as in eq.\eqref{eq:thermal_oracle}, here we will investigate the no. of queries and samples required to learn whether $f$ is balanced or constant using different heat exchanges as depicted in Fig.~\ref{fig:query_model}. (i) For an input $n$-bit string $x$ a simple heat exchange $\mathtt{SWAP}_{S,x} = \mathtt{SWAP} \otimes \left(\bigotimes_{y \in \{0,1\}^{\times n} \setminus x} \mathbb{1}\right)$ acts on $\rho_S(\beta) \otimes \tau_f$ to swap the probe qubit $\rho_S$ for qubit $\tau_x$ in the machine. The agent can then collect samples of each $\tau_x$ and carry out energy measurements for each collection, solving the problem with finite error given $2^{n-1} + 1$ heat exchanges/queries and a constant no. of samples per query \footnote{We will examine the no. of samples in the \textit{Reading Out} section.}. (ii) A classical mixture over all $n$-bit strings $\mathbb{1}/2^n = 1/2^n \sum_{x\in \{0,1\}^{n}}\ketbra{x}{x}$ can instead be input resulting in a probabilistic $\mathtt{SWAP}_{S,x}$ over every possible input in a single query. Letting $\tau_1 =1/(1 + e^{-\beta_M E_1})\left(\ketbra{0}{0} + e^{-\beta_M E_1}\ketbra{1}{1}\right)$ denote $\tau_x$ in the case where $f(x) = 1$ and similarly $\tau_2$ for the case that $f(x) = 0$ we see that $\tau_f$ can take one of three forms $\tau_f \in \left\{\tau_1^{\otimes N}, \tau_2^{\otimes N}, \pi\left(\tau_1^{\otimes N/2}\otimes \tau_2^{N/2}\right)\right\}$, where $\pi \in \mathcal{S}_N$ is permutation for different balanced functions. As such, the result of $\mathtt{SWAP}_{x,S}$ on $\rho_S \otimes \tau_f$ conditioned on $\mathbb{1}/2^n$ is to return the probe in the state $\tau_1$ or $\tau_2$ in the constant cases and $(\tau_1 + \tau_2)/2$ for any balanced case. In a single query we obtain the desired information but we require a constant number of samples for an error $\delta$ of $\Theta(\log(1/\delta)/D(\tau_1||(\tau_1~+~\tau_2)/ 2))$. At best in this model, i.e. for $\tau_1 = \ketbra{0}{0}, \tau_2=\mathbb{1}/2,$ this quantity $\Theta(\log(1/\delta)/\log (4/3))$ is slightly worse than classical random sampling~\cite{sup_mat}. Classical randomness and bipartite quantum operations
on thermal states fail to beat classical probabilistic computation, can stronger thermodynamic operations do better?

\textit{Thermal Kickback} We will now and for the remainder of the manuscript consider a stronger heat exchange between the system and the oracle's thermal machine i.e. the largest change in population in $S$ given by a single energy level exchange---which can for some parameters beat random sampling. Fig~\ref{fig:query_model} (iii) consider a vector valued function $X(y_0,y_1, \dots, y_{2^n})$ with $2^{n}$ entries, each corresponding to an $n$-bit string, which maps the vector of strings to a $2^n$-bit string where if $x_i$ was present in the input vector it is mapped to a $1$ in the $2^n$-bit string and 0 if not e.g. $(x_0, x_1) \,\,\, \longrightarrow \,\,\, 110\dots0$. The heat exchange $V(X) = \ketbra{0_S X}{1_S X~\oplus~1} + \ketbra{1_S X \oplus 1}{0_S X} + \mathbb{1}_\text{Rest}$ is then carried out, where the energy level $\ket{0_S X}$ featuring no excitations in the probe and excitations in the thermal machine oracle related to input bit strings $x_i$ represented by 1s in $X$, is exchanged with its Hamming weight conjugate $\ket{1_S X \oplus 1}$ featuring opposite excitations. The case where all $n$-bit strings are input i.e. $X= 1^{N}$ corresponds to the \textit{virtual qubit subspace swap}~\cite{skrzypczyk_10,brunner_12,ralph_swap} a mechanism often considered in quantum thermodynamics as it is the optimal heat exchange for asymptotic cooling~\cite{fab_limits} and has been used in the design of quantum heat engines, clocks, thermometers and recently artificial neuron circuits~\cite{skrzypczyk_10,erker_17,hofer_17,bartosik2024thermodynamicalgorithmsquadraticprogramming}. We will focus on $X = 1^{N}$ as input as it will output global information about $f(x)$. The interaction $V(1^N)$ between a probe with gap $\omega$ and inverse temperature $\beta_S$ and a thermal machine oracle at temperature $\beta_M$ exchanges $\ket{0_S 1^N}$ with population $e^{-\beta_M |\Gamma|}/\mathcal{Z}_S\mathcal{Z}_f$ and $\ket{1_S 0^N}$ with population $e^{-\beta_S \omega}/\mathcal{Z}_S\mathcal{Z}_f$ leads to a ground state population (g.s.p.) in $S$ of
\begin{align}
    p_0' = \bra{0}\rho'_S\ket{0} = \frac{1 + \mathcal{Z}^{-1}_f(e^{-\beta_S \omega} - e^{-\beta_M |\Gamma|})}{\mathcal{Z}_S},\label{eq:pop_after}
\end{align}
where $|\Gamma| = \text{tr}\{\Gamma\}$. Recalling that the g.s.p. of a qubit need take the form $p_0' = 1/(1 + e^{-\beta'_S \omega})$ where $\beta'_S = 1/T'_S$ is the inverse temp. after heat is exchange we can solve for $\beta'_S = \omega^{-1}\log\left(\frac{p'_0}{1-p'_0}\right)$ to find that 
\begin{gather}
   \beta'_S = \frac{1}{\omega} \log\left(\frac{1 + \mathcal{Z}^{-1}_f (e^{-\beta_S \omega} - e^{-\beta_M |\Gamma|})}{e^{-\beta_S \omega} - \mathcal{Z}^{-1}_f(e^{-\beta_S\omega} - e^{-\beta_M |\Gamma|})}\right).\label{eq:dj_temp}
\end{gather}
as detailed in the End Matter, where $|\Gamma| = N E_1$ or $NE_2$ if $f(x)$ is constant and $|\Gamma| = N/2(E_1 + E_2)$ if $f(x)$ is balanced allowing us to solve the Deutsch-Jozsa Problem by measuring the temperature of a qubit after a single heat exchange. The sample complexity of this measurement is investigated in the coming paragraphs where for some parameters, advantage over random sampling is possible. In the End Matter, expressions for the g.s.p. and temperature of the probe after query for general input $X$ that is not $X = 1^N$ is provided.

\textbf{A Linear Thermal Oracle -- Hamming Weight Detection of Secret Strings} In the Bernstein - Vazirani problem~\cite{bernstein_vazirani} an agent is given access to an oracle encoding a Boolean function $f(x) : \{0,1\}^{\times n} \rightarrow \{0,1\}$ with the promise that $f(x)$ is formed by taking the dot product under $\mod 2$ with a \textit{secret} $n$-bit string $s$, that is $f(x) = x\cdot s = x_1 s_1 \oplus x_2 s_2  \oplus \dots \oplus x_ns_n$. The solution of this problem is to determine $s$ by querying a unitary oracle encoding $f(x)$ as in the Deutsch-Jozsa problem. A restricted version of this problem, \textit{Hamming weight detection}, is also solvable given a single heat exchange with a qubit when encoded in a thermal machine provided an energy gap promise, this time with linearly many qubits. Let the oracle construct an $n$ thermal qubit machine at temp. $T_M$ such that each qubit has an energetic gap $s_i\gamma$ and state $\tau_i = \frac{1}{(1+e^{-\beta_M s_i\gamma})}\left(\ketbra{0}{0} + e^{-\beta_M s_i\gamma}\ketbra{1}{1}\right) \, : s_i \in \{0,1\}$ where $s_i$ is the $i$th secret bit, $\gamma$ is a fixed energy gap and the global state is $\tau_f = \bigotimes^n_{i = 1} \tau_i$. The $2^n$ energy levels of $\tau_f$, $\ket{x} : x \in \{0,1\}^{\times n}$ have populations $e^{-\beta_M s \cdot x \cdot \vec{\gamma}}$ where $s \cdot x \cdot \vec{\gamma} = s_1x_1\gamma + s_2x_2\gamma + \dots s_{n}x_n\gamma$ and $x_i \in \{0,1\}$ denotes whether the $i$th qubit is excited in $\ket{x}$. Thus each population of $\tau_f$ features the output $f(x) = s\cdot x$ and be queried using the probe via the thermal kickback introduced before $V(x) = \ketbra{0_S \, x}{1_S\, x \oplus 1} + \ketbra{1_S \, x \oplus 1}{0_S \, x} + \mathbb{1}_\text{Rest}$. Notably, for $x = 1^n$ we have the post-query probe population $p'_0 = \left(1 + \mathcal{Z}_f^{-1}\left(e^{-\beta_S \omega} - e^{-\beta_M \#(s) \gamma}\right)\right)\mathcal{Z}^{-1}_S,$ 
where $\#(s)$ is the Hamming weight or no. of 1s in $s$ allowing an agent to detect $\#(s)$ with knowledge of $\gamma$. This presents a challenge we did not encounter in the DJ problem, in this scenario the no. of temperatures we need to distinguish to determine $\#(s)$ scales linearly with the size of $s$ posing a problem of distinguishability. This translates this computational problem into one of metrology~\cite{Mehboudi_2019} or multi-hypothesis testing which we leave for future work. If the oracle used different energies $\gamma_i$ for each $\tau_i$ this exchange could also discriminate strings but would stricter promise condition where the agent knows each $\gamma_i$ for read out.

\textbf{Thermodynamic Cost of a Query} 
Given the nature of the model, we are able to investigate the energetics and thermodynamics of querying. (i) The change in ground state population in $S$ induced by $V(X)$ can be positive or negative, heating or cooling the probe. From eq.\ref{eq:pop_after} we see that for cooling $\Delta p_0 = p_0' - p_0 > 0$ if $e^{-\beta_S \omega} - e^{-\beta_M|\Gamma|}>0$ and similarly for heating $\Delta p_0 < 0$ if $e^{-\beta_S \omega} - e^{-\beta_M|\Gamma|}>0$. This provides two regimes where the probe is 
\begin{gather}
    \text{cooled : } \frac{\omega}{T_S} < \frac{|\Gamma|}{T_M}, \hspace{0.5cm}\text{or}\hspace{0.5cm}     \text{heated : } \frac{\omega}{T_S} > \frac{|\Gamma|}{T_M}, \label{eq:conditions}
\end{gather}
by the query depending on the ratios of energy and temperature of the probe $\omega/T_S$ and the ratio of sum of all energy gaps and temperature $|\Gamma|/T_M$ of the thermal machine oracle.
ii) $e^{-\beta_M|\Gamma|}$ could become too small in comparison to $e^{-\beta_S \omega}$ making the probe energetically insensitive to this heat exchange and so, the properties of $f(x)$. For ii) we would like $|\Delta p_0| > c$ where $ 0\!~<~\!c\!~<\!~1 - p_0$ is a sensitivity constant. This imposes the condition $c<|e^{-\beta_S\omega} - e^{-\beta_M|\Gamma|}|/\mathcal{Z}_f\mathcal{Z}_S$, which leads to tightened versions of the conditions eq.\eqref{eq:conditions}. Focusing on cooling, we require $e^{-\beta_S \omega} -e^{-\beta|\Gamma|}>c\mathcal{Z}_S\mathcal{Z}_f$ so that rearranging and taking logarithms gives the inequality $\omega/T_S < - \log\left(c\mathcal{Z}_S\mathcal{Z}_f + e^{-\beta_M |\Gamma|}\right)$ where the r.h.s is positive if $0 < c\mathcal{Z}_S\mathcal{Z}_f + e^{-\beta_M |\Gamma|} \leq 1$. Lastly, for the temperature of the probe after query to be well-defined we examine eq.\eqref{eq:dj_temp} which can be simplified as $\beta'_S = \omega^{-1}\log\left((1+\mathcal{Z}_S\Delta p_0)/(e^{-\beta_S \omega}-\mathcal{Z}_S\Delta p_0)\right)$. In the case of cooling we require $e^{-\beta_S \omega}> \mathcal{Z}_S\Delta p_0$ for heating $1 > \mathcal{Z}_S\Delta p_0$ for the term in the logarithm to be positive. The inequalities considered here together with their derivation are given in full in the supplemental material~\cite{sup_mat}.

The probe and machine qubits are embedded in an environment at temperature $T_M$ where the probe is pushed out of equilibrium to $T_S$ in preparation for query. Resetting the machine after query via thermalisation would cost $E_\text{Diss} = \Delta p_0 |\Gamma|$ in dissipation, whilst setting the probe out of equilibrium again $E_\text{reset} = \Delta p_0 \omega$ of work.

\textbf{Reading Out} Encoding the solution of a decision problem in the temperature of a thermal qubit presents two related difficulties. (i) The temperatures corresponding to different outcomes of the decision problem must be physically distinguishable. (ii) The number of samples of the probe required to estimate its temperature, and so learn the outcome of the problem, should be small in the size of the problem. Ideally, not exceeding the sample complexity of a classical probabilistic solution, suggesting an advantage. We examine here these difficulties in the context of reading out the solution to a Deutsch-Jozsa problem via thermal kick back. Consider the g.s.p. of the probe after thermal kick back for the different outcomes $p^\text{Bal}_0,p^\text{Const$_1$}_0,p^\text{Const$_2$}_0 $ and w.l.o.g. let $p^\text{Const}_0 = p^\text{Const$_1$}_0$ corresponding to $|\Gamma_\text{Const}| = N E_1$, then we are interested in $p^\text{Const}_0 - p^\text{Bal}_0 > t$ which is positive for $E_1 > E_2$ so that $NE_1 > N/2(E_1 +E_2)$. Since $p'_0 = (1 + \mathcal{Z}_S \Delta p_0)/\mathcal{Z}_S$ the inequality of interest becomes $|\Delta {p_0}^\text{Const} - \Delta p_0^\text{Bal}|> t$. We will consider the case where the query cools the probe so that $\Delta {p_0}^\text{Const},\Delta p_0^\text{Bal} > 0$. We then have
$\frac{e^{-\beta_S \omega} - e^{-\beta_M |\Gamma_\text{Const}|}}{\mathcal{Z}^\text{Const}_f\mathcal{Z}_S} - \frac{e^{-\beta_S \omega} - e^{-\beta_M |\Gamma_\text{Bal}|}}{\mathcal{Z}^\text{Bal}_f\mathcal{Z}_S} > t,$
in general. The maximal cooling is possible when the probe is initially in the maximally mixed state where $e^{-\beta_S\omega} = 1$ and $\mathcal{Z}_S$. In this setting the above condition simplifies to a constraint on the difference between the largest population of the oracle's machine in the two cases\begin{gather}
\frac{1}{\mathcal{Z}^\text{Const}_f} - \frac{1}{\mathcal{Z}^\text{Bal}_f} > 2t,  \label{eq:dist_cond}
\end{gather}
as detailed in the End Matter, where we also provide a plot showing values of $t$ reachable for different $E_1$, $E_2$ and input size $n$.

Determining whether $f(x)$ is balanced or constant reduces to a \textit{classical} binary hypothesis testing scenario~\cite{thomas_cover}. Here, with access to samples $\tau'_S$ of the post interaction probe and energy basis measurements, an agent must determine whether $S'$ is (Hypothesis A) in temperature $T_\text{Bal}$ or (Hypothesis B) not in $T_\text{Bal}$ i.e. it is in temperature $T_\text{Const$_1$}$ or $T_\text{Const$_2$}$. This is a classical hypothesis test since $\tau'_S(T_\text{Bal})$ and $T_\text{Const.}$ are diagonal in the energy eigenbasis and so commute. As long as the type A false positive error $\delta$ can be controlled such that $0 < \delta < 1$ then the Chernoff-Stein Lemma~\cite{thomas_cover} states that the no. of samples $n^*$ of $\tau'_S$ required to determine the scenario are lower bounded by $\log(1/\delta)/D(\tau'_S(T_\text{Bal})||\tau'_S(T_\text{Const.}))$ where w.l.o.g $\tau'_S(T_\text{Const.})$ corresponds to the temp. $T_\text{const$_1$}$ or $T_\text{const$_2$}$ which minimises the relative entropy $D(\tau'_S(T_\text{Bal})||\tau'_S(T_\text{Const.}))$. Using Pinsker's Inequality~\cite{thomas_cover} $D(\tau'_S(T_\text{Bal})||\tau'_S(T_\text{Const.})) > 2d_{TV}(\tau'_S(T_\text{Bal}),\tau'_S(T_\text{Const.}))^2$ where $d_{TV}(\cdot,\cdot)$ is the total variation distance and since these are qubit thermal states and so binary distributions, we have $d_{TV}(\tau'_S(T_\text{Bal}),\tau'_S(T_\text{Const.})) = |~p^\text{Bal}_0~-~p^\text{Const}_0~|$. When the probe satisfies the distinguishability condition eq.\eqref{eq:dist_cond} we thus have the lower bound 
\begin{gather}
    n^* > \frac{\log(1/\delta)}{2t^2} \label{eq:samp_bound}
\end{gather}
which is independent of the problem size! Given that this is a classical binary hypothesis test, this lower bound is asymptotically achievable using a likelihood ratio test~\cite{thomas_cover}.

\textbf{Discussion \& Conclusion} As long as one can energetically ensure the distinguishability condition eq.\eqref{eq:dist_cond}, the number of samples of $S'$ required to solve the Deutsch-Jozsa problem is independent of $n$, e.g. letting $\delta = t = 0.1$ one requires at least $50\log(10) \approx 116$ samples for any $n$ using the bound eq.\eqref{eq:samp_bound}. This suggests that while the quantum solution requires one query and one measurement, and the classical solution requires $2^{n-1} + 1$ queries, quantum thermodynamics offers an intermediate regime requiring 1 query and a constant no. of samples $\Theta\left(\frac{\log(1/\delta)}{2t^2}\right)$. Whilst not advantageous over classical approaches for small $n$, this appears to offer advantage for intermediate to large $n$. In particular, for the parameters $\delta = t = 0.1$ the sample complexity is lower than the classical query complexity for deterministic solutions at $n > 8$. 
If one considers probabilistic solutions to the Deutsch-Jozsa problem which allow for false positive errors, there is no longer an exponential separation between classical and quantum approaches~\cite{Nielsen_Chuang_2010}. We examine how our sample complexity compares to that of a probabilistic classical approach in the supplemental material and find proof-of-principle advantage dependent on the size of $t$.

In this context, there are two ways to interpret having access to multiple samples of the post-interaction probe. Firstly, since $\tau'_S$ is diagonal in the energy eigenbasis, a single sample may be used to unitarily reconstruct $\tau'_S$ in $n^*$ pure ancillas (e.g. using a \texttt{CNOT} conditioned on $S$ acting on an ancilla) allowing for the desired repeated measurements. Secondly, the probe and machine can be thermodynamically reset as described earlier, allowing for repeated sampling. 

One might wonder where the \textit{quantumness} in our model is since no coherences were generated. Indeed, under which parameters autonomous thermal machines are genuinely quantum i.e. their dynamics cannot be described by a purely stochastic master equation is an active research question~\cite{almanzamarrero2025certifyingquantumenhancementsthermal}. An interesting direction for future work would be to relate the distinguishability $t$ to the quantumness of the model, thereby showing that sample complexity advantages stem from genuinely quantum effects.

In this work we examined an agent's ability to learn global properties of Boolean functions encoded in the energetic structure of a thermal machine via a heat exchange with a qubit. This serves as a fascinating model for the physics of quantum query complexity and a provocation to think of alternative models of quantum computation. In the End Matter we provide a sketch of an experimental implementation of the ideas discussed.

\textbf{Acknowledgments} The author thanks  Marcus Huber, Ale de Oliveira Jr., Florian Meier, Federico Fedele, Pharnam Bakhshinezhad, Moha Mehboudi \& Gabriel Landi for fruitful discussion and Ben Stratton for many useful conversations on the ideas in this manuscript. We acknowledge funding from the European Research Council (Consolidator grant ‘Cocoquest’ 101043705) and Steve Campbell for his hospitality and support during a visit to University College Dublin where parts of this work were completed. The code for generating the presented plots is available at~\cite{code}.

\providecommand{\noopsort}[1]{}\providecommand{\singleletter}[1]{#1}%

\appendix

\section{End Matter}
\paragraph*{Thermal Kickback Temperature}
\label{sec:djderiv}To obtain the probe qubit temperature eq.\eqref{eq:dj_temp} after the query, which determines whether $f(x)$ is constant or balanced let's consider the following. First, the ground state population of a thermal qubit with Hamiltonian $H = 0\ketbra{0}{0} + \omega\ketbra{1}{1}$ at inverse temperature $\beta$ can be expressed as $p_0 = \frac{1}{1+ e^{-\beta \omega}}$. Then solving for the inverse temp. $\beta$ we obtain $
    \beta = \frac{1}{\omega}\log\left(p_0/1-p_0\right). $
Next, we found the ground state population of the probe after the query eq.\eqref{eq:pop_after} to be $p_0' = \frac{1 + \mathcal{Z}^{-1}_f(e^{-\beta_S \omega} - e^{-\beta_M |\Gamma|})}{\mathcal{Z}_S}$ and note that the change in ground state population is therefore
\begin{align}
    \Delta p_0 &= p'_0 - p_0\\
    &= \frac{1 + \mathcal{Z}^{-1}_f(e^{-\beta_S \omega} - e^{-\beta_M |\Gamma|})}{\mathcal{Z}_S} - \frac{1}{\mathcal{Z}_S}\\
    &= \frac{e^{-\beta_S \omega} - e^{-\beta_M |\Gamma|}}{\mathcal{Z}_f\mathcal{Z}_S}, \label{eq:delta}
\end{align}
so that $\mathcal{Z}^{-1}_f(e^{-\beta_S \omega} - e^{-\beta_M |\Gamma|}) = \mathcal{Z}_S \Delta p_0$. Finally, to obtain $\beta'_S$ the inverse temperature of the probe qubit after querying the thermal machine oracle we substitute $p'_0$ into the logarithmic expression for $\beta$ given earlier to obtain $\log\left(\frac{p_0'}{1 - p'_0}\right) = \log(p'_0) - \log(1 - p'_0)$
\begin{align}
    &= \log\left(\frac{1 + \mathcal{Z}_S \Delta p_0}{\mathcal{Z}_S}\right) - \log\left(\frac{\mathcal{Z}_S - 1- \mathcal{Z}_S \Delta p_0}{\mathcal{Z}_S}\right)\\
    &= \log(\mathcal{Z}_S) - \log(\mathcal{Z}_S) + \log(1+\mathcal{Z}_S\Delta p_0) \nonumber \\ 
    &\hspace{3.8cm}- \log(\mathcal{Z}_S - 1 - \mathcal{Z}_S\Delta p_0)\\
    &= \log\left(\frac{1 + \mathcal{Z}^{-1}_f (e^{-\beta_S \omega} - e^{-\beta_M |\Gamma|})}{\mathcal{Z_S} - \mathcal{Z}^{-1}_f(e^{-\beta_S\omega} - e^{-\beta_M |\Gamma|}) -1}\right)
\end{align}
where we find
\begin{align}
    \beta'_S = \frac{1}{\omega} \log\left(\frac{1 + \mathcal{Z}^{-1}_f (e^{-\beta_S \omega} - e^{-\beta_M |\Gamma|})}{\mathcal{Z_S} - \mathcal{Z}^{-1}_f(e^{-\beta_S\omega} - e^{-\beta_M |\Gamma|}) -1}\right)
\end{align}
For a general $V(X)$ as described in the main text, the change in temperature of the probe is instead
\begin{align}
 \hspace{-1cm}   \beta'_S (X) &= \frac{1}{\omega} \log\left(\frac{1 + \mathcal{Z}^{-1}_f (e^{-\beta_S \omega - \beta_M (|\Gamma|- X \cdot \Gamma)} - e^{-\beta_M X\cdot \Gamma})}{e^{-\beta_S \omega} - \mathcal{Z}^{-1}_f(e^{-\beta_S \omega - \beta_M (|\Gamma|- X \cdot \Gamma)} - e^{-\beta_M X\cdot \Gamma}) }\right) \nonumber \\
    &= \frac{1}{\omega} \log\left(\frac{1 + \mathcal{Z}_S\Delta p_0 (X)}{e^{-\beta_S \omega} - \mathcal{Z}_S \Delta p_0 (X)}\right).
\end{align}

\begin{figure}[h]
    \centering
    \includegraphics[width=\linewidth]{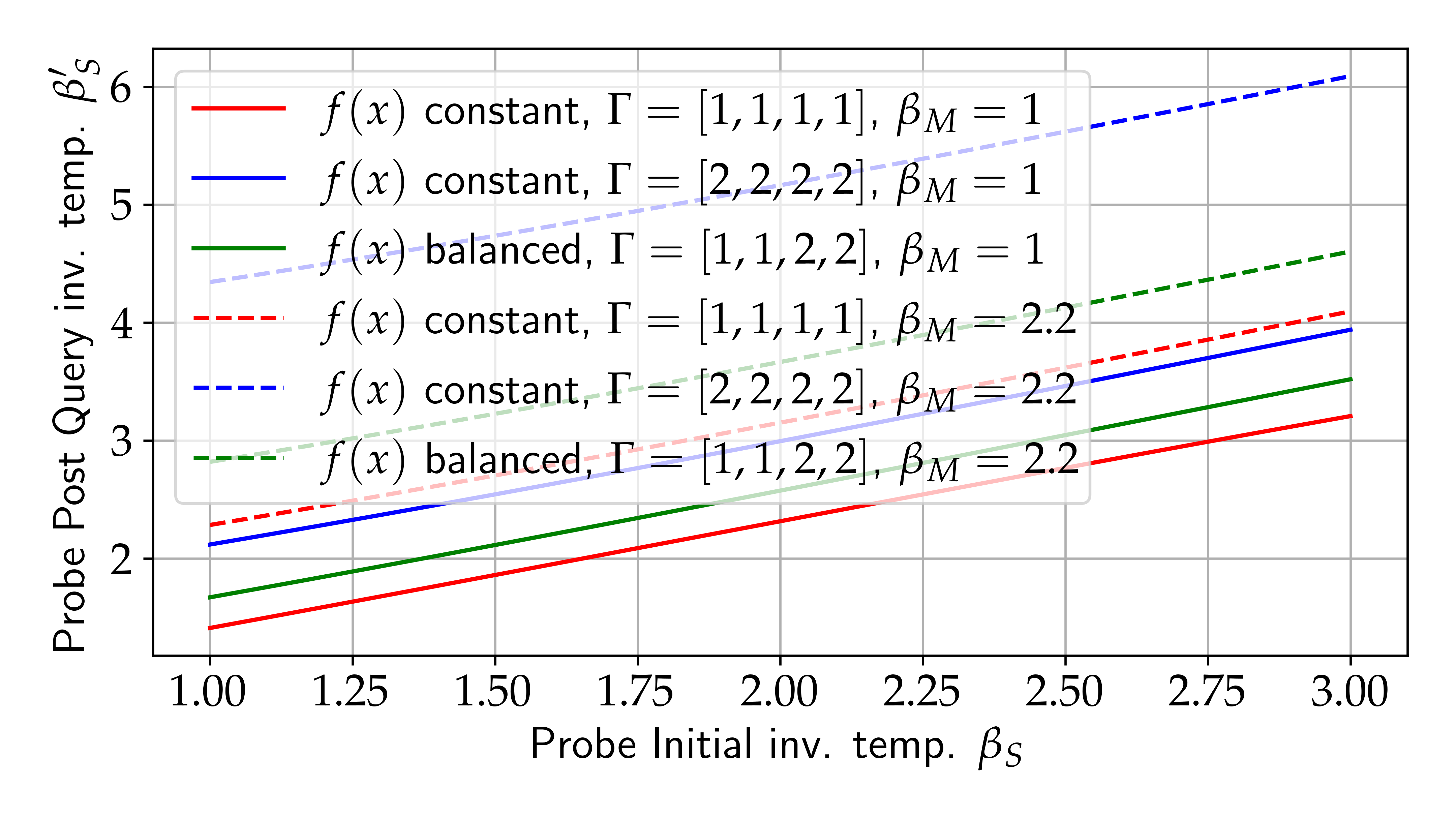}
    \caption{A plot showing the inv. temp. of the probe with $\omega = 1$ after thermal query $\beta'_S$ against the initial temp. $\beta_S$ of the probe for the 2-bit DJ problem. We see that the different outcomes are less distinguishable as the thermal oracle qubits become warmer i.e. smaller $\beta_M$ (solid lines).}
    \label{fig:dj_temps}
\end{figure}

\paragraph*{Probe Conditions for Distinguishability}
\label{sec:disting}

For distinguishability we require that the difference between the  ground state population of the probe after interaction for different cases be larger than a threshold value $t$. In the Deutsch-Jozsa case without loss of generality let $p^\text{Const}_0 = p^\text{Const$_1$}_0$ corresponding to $|\Gamma_\text{Const}| = N E_1$, then we are interested in 
\begin{gather}
	p^\text{Const}_0 - p^\text{Bal}_0 > t
\end{gather}
which is positive for $E_1 > E_2$ so that $NE_1 > N/2(E_1 +E_2)$. Substituting in the form of the ground state population after thermal kickback eq.~\eqref{eq:pop_after} we have
\begin{align}
   &\frac{1 + \mathcal{Z}_S{\Delta p_0}^{\text{Const}}}{\mathcal{Z}_S} - \frac{1 + \mathcal{Z}_S{\Delta p_0}^{\text{Bal}}}{\mathcal{Z}_S} > t\nonumber \\
   &{\Delta p_0}^{\text{Const}} - {\Delta p_0}^{\text{Bal}} > t.
\end{align}
We will consider the case where the query cools the probe so that $\Delta p_0 > 0$ for both scenarios. Expanding using eq.\eqref{eq:delta} we have
\begin{align}
&\frac{e^{-\beta_S \omega} - e^{-\beta_M |\Gamma_\text{Const}|}}{\mathcal{Z}^\text{Const}_f\mathcal{Z}_S} - \frac{e^{-\beta_S \omega} - e^{-\beta_M |\Gamma_\text{Bal}|}}{\mathcal{Z}^\text{Bal}_f\mathcal{Z}_S} > t\\
&e^{-\beta_S\omega}(\mathcal{Z}^\text{Bal}_f - \mathcal{Z}^\text{Const}_f) - \chi > t \mathcal{Z}_S \mathcal{Z}^\text{Bal}_f\mathcal{Z}^\text{Const}_f
\end{align}
where we have cross-multiplied to obtain a common denominator, then multiplied both sides by the common denominator $\mathcal{Z}_S \mathcal{Z}^\text{Bal}_f\mathcal{Z}^\text{Const}_f$ and adopted the notation $\chi = \mathcal{Z}^\text{Const}_f e^{-\beta_M |\Gamma_\text{Bal}|} - \mathcal{Z}^\text{Bal}_f e^{-\beta_M |\Gamma_\text{Const}|}$. The strongest scenario for cooling involves the probe initially in a maximally mixed state which sets $e^{-\beta_S \omega} = 1$ and $\mathcal{Z}_S = 2$ in this inequality. Also note that $\chi$ is exponentially suppressed and can be omitted up to additive error $\mathcal{O}(\chi)$ giving the condition $\mathcal{Z}^\text{Bal}_f - \mathcal{Z}^\text{Const}_f > 2t\mathcal{Z}^\text{Bal}_f\mathcal{Z}^\text{Const}_f$, which leads to
\begin{align}
\frac{\mathcal{Z}^\text{Bal}_f - \mathcal{Z}^\text{Const}_f}{\mathcal{Z}^\text{Bal}_f\mathcal{Z}^\text{Const}_f} > 2t \iff \frac{1}{\mathcal{Z}^\text{Const}_f} - \frac{1}{\mathcal{Z}^\text{Bal}_f} > 2t. 
\end{align}
Note that $1/\mathcal{Z}_f$ is the ground state population of the thermal machine oracle, thus this inequality communicates that a difference of $t$ in ground state populations of the probe for different cases is only possible if this corresponds to a difference of $2t$ in machine g.s.p.s for these two cases.
\begin{figure}
    \centering
    \includegraphics[width=\linewidth]{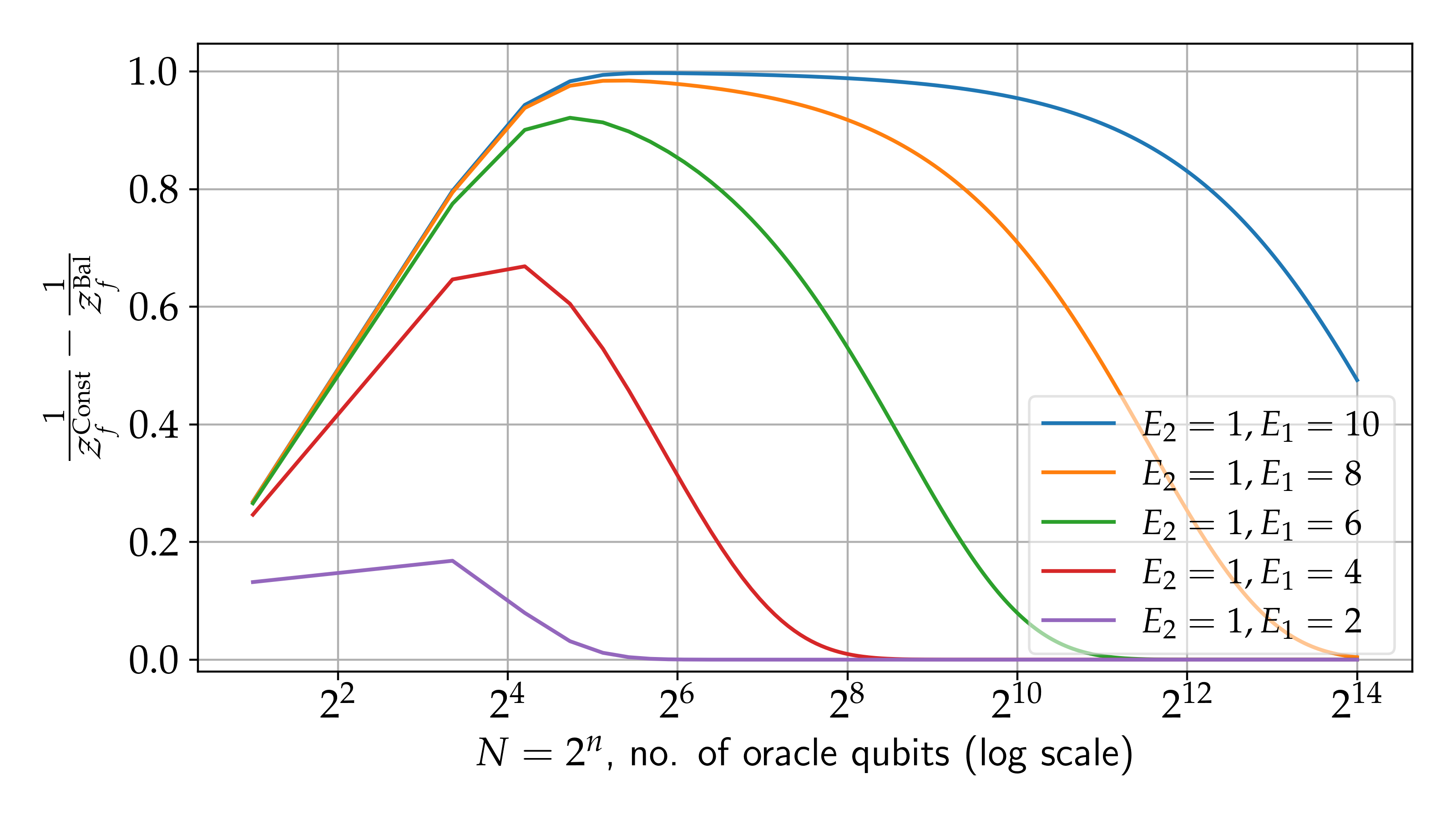}
    \caption{A plot showing the L.H.S of inequality eq.\eqref{eq:dist_cond}\\ $(1 + e^{-\beta_M E_1})^{-N} - (1 + e^{-\beta_M E_1})^{-N/2}(1 + e^{-\beta_M E_2})^{-N/2}$ for $\beta_M = 1$ and various oracle qubit energies $E_1$ and $E_2$, showing that $t \in [0,0.5]$ is achievable with different energies for different $n$.}
    \label{fig:placeholder}
\end{figure}

\paragraph*{Experimental Implementation}
\label{sec:detuning}

As a stimulating example of the experimental feasibility of our ideas, consider that an implementation of the 3 Bit Bernstein - Vazirani Oracle presented can be implemented in a quantum thermodynamics experiment~\cite{Pekola2015}, using a quantum dot~\cite{rmp_trans_02,rmp_spin_dot_07,dots_rev_mod} or transmon qubit~\cite{rmp_transmon,Aamir2025} set up. Here three quantum dots or transmon qubits $M_1, M_2, M_3$ are each prepared in an individual thermal state at the same temperature $T_M$ with gate voltages or radio frequency (RF) pulses respectively being used to modulate their energetic gaps $\Delta_1(s_1), \Delta_2(s_2)$ and $\Delta_3(s_3)$ such that $\Delta_i(s_i) = s_i \gamma_i$ with $\gamma_i$ being a fixed tunable energy parameter e.g. gate voltage or RF pulse for qubit $M_i$ and $s_i \in \{\epsilon_1, \epsilon_2\}$ being an energetic bias which encodes the unknown string. A random number generator can then be used to shift the energetic gaps by $s_1, s_2$ and $s_3$. A fourth probe qubit is prepared in a thermal state at temp. $T_S > T_M$ with gap $\omega$, with the goal of estimating $s$ by exchanging heat with $M$. The machine gaps $\gamma_1, \gamma_2, \gamma_3$ would need to be engineered and $\epsilon_1$ and $\epsilon_2$ small enough such that $\omega \approx \Delta_1(s_1) + \Delta_2(s_2) + \Delta_3(s_3)$, for all possible strings $s_1s_2s_3 \in \{0,1\}^{\times 3}$, ensuring that an energy selective coherent Rabi flip-flop interaction can occur between the $\ket{0111}$ and $\ket{1000}$ energy levels via the interaction Hamiltonian
\begin{gather}
    H_\text{query} = g(\sigma^{(S)}_{-}\sigma^{(M_1)}_{+}\sigma^{(M_2)}_{+}\sigma^{(M_3)}_{+} + h.c.)
\end{gather}
which will effectively generate the desired $U_\text{query}$. The gap structure of the three oracle qubits is then estimated from a temperature reading of the probe qubit $S$ to determine the secret string. This presents the challenge of such an experiment implementing a choice of parameters for which $\epsilon_1$ and $\epsilon_2$ are small enough to allow enough coherent heat exchange (minimising detuning) but not so small that different strings are no longer thermodynamically distinguishable. In particular we would like the energetic condition $\omega = s_1 \gamma_1 + s_2 \gamma_2 + s_3 \gamma_3$ be satisfied for any secret string $s_1s_2s_3 \in \{\epsilon_1,\epsilon_2\}^{\times 3}$. This is clearly not possible in general, so let's examine what happens if one approximately relaxes this condition introducing some detuning. \textit{How would this imperfection influence the distinguishability of the temperatures corresponding to different strings?} Let's assume that these bias energies take the form $\epsilon_1 = 1 - \epsilon, \epsilon_2 = 1 + \epsilon$ for a small $\epsilon > 0$ then the total oracle energy deviates slightly from $\omega$, producing a detuning for a given string $s$, $\delta(s) = s_1 \gamma_1 + s_2 \gamma_2 + s_3 \gamma_3 - \omega$. To keep the coherent Rabi oscillation strength large, the detuning must satisfy $\delta(s) \ll g$, where $g$ is the interaction strength. In this regime, the flip-flop probability is given by
\begin{gather}
P_\text{flip}(t) \approx \frac{g^2}{g^2 + \delta^2} \sin^2\left(\frac{1}{2}\sqrt{g^2 + \delta^2}\,t\right),
\end{gather}
indicating that population transfer remains high for small detuning. For short times $t$ this simplifies to $\eta(\delta) = \frac{g^2}{g^2 + \delta^2}$ which acts as a suppression factor in the change in population inverse temperature of the probe system  

\begin{equation}
\hspace{-0.5cm}\beta_S'(\!\delta\!) \!=\! \frac{1}{\omega} \log\!\!\left(\!\! \frac{1 + \eta(\delta) \, \mathcal{Z}^{-1}_f (e^{-\beta_S \omega} - e^{-\beta_M |\Gamma|})}{\mathcal{Z}_f - \eta(\delta) \, \mathcal{Z}^{-1}_f (e^{-\beta_S \omega} - e^{-\beta_M |\Gamma|}) - 1}\!\! \right)\!.
\end{equation}

\begin{figure}[h]
    \centering
    \includegraphics[width=\linewidth]{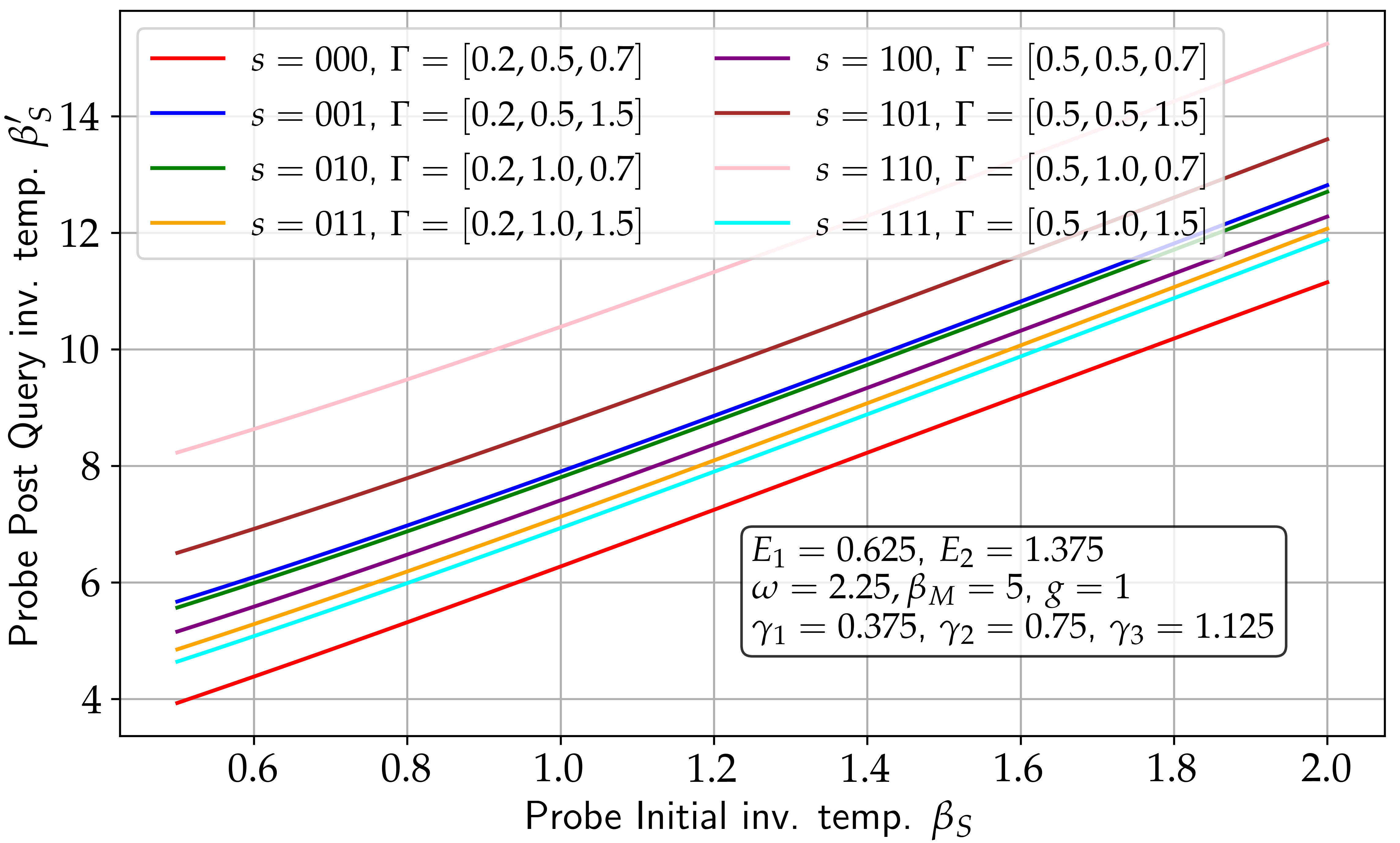}
    \caption{A plot showing the inv. temp. of the probe after thermal query $\beta'_S$ against the initial temp. $\beta_S$ of the probe for the three bit Bernstein-Vazirani problem i.e 3 qubit oracle in the presented experimental implementation. In this figure we see that each of the 8 potential secret strings is distinguishable in the temp. of the probe after the heat exchange despite the detuning condition $\delta(s)$.}
    \label{fig:detuning}
\end{figure}

We plot the inv. temperature of the probe after querying the oracle in this realisation of the 3 bit Bernstein-Vazirani example in Fig.~\ref{fig:detuning}.
experimental
\clearpage
\onecolumngrid
\appendix
\section{Supplemental Material}
\subsection{The Oracle prepares a Thermal Machine} 
In this section we will consider how an oracle with the ability to call a Boolean function $f(x)$ for a given input $x$ could construct a thermal machine as described in the main text. Let us consider a channel $\mathcal{T}(\cdot)$ which takes as input an $n$-bit string x and a single qubit pure state $\ket{0}$ where $x \in \{0,1\}^{\times n}$ encodes the input for which the oracle is querying the function and $\ket{0}$ is qubit initialised in a pure state which will come to form a part of the machine. The channel $\mathcal{T}(\cdot)$ will be formed of Kraus operators 
\begin{gather}
    K_x = \ketbra{x}{x} \otimes \left( (1 - p_x)\mathbb{1} + p_x \sigma_x\right) 
\end{gather}
which are conditional amplitude damping or thermalisation channels. In particular, for a given input $x$ the oracle computes $E_x = f(x) + E$ where $E > 0$ is an arbitrary fixed positive constant,  by querying $f(x)$ and obtains the probability $p_x = e^{-\beta_M E_x}/(1 + e^{-\beta_M E_x})$ which defines the parameter in the amplitude damping channel. From here we see that the oracle will be able to obtain the thermal qubit 
\begin{gather}
    \tau_x = \frac{1}{1 + e^{-\beta_M E_x}}\left(\ketbra{0}{0} +e^{-\beta_M E_x}\ketbra{1}{1} \right),
\end{gather}
for each input string $x \in \{0,1\}^{\times n}$ by applying $\mathcal{T}(\ketbra{x}{x}\otimes \ketbra{0}{0})$ allowing them to construct the thermal machine described in the main text with access to an $n$-bit classical register and $2^n$ qubits which will be used to generate the thermal machine.

\subsection{Thermodynamic Cost of a Query} We present here for completeness, the derivations of the inequalities discussed in the \textbf{Thermodynamic Cost of a Query} Section of the main text in full. Recall as presented in the End Matter that 
\begin{align}
    \Delta p_0 &= p_0 ' -p_0 \\
               &= \frac{e^{-\beta_S\omega} - e^{-\beta_M |\Gamma|}}{\mathcal{Z}_S \mathcal{Z}_f}.
\end{align}
If the query cools the probe then $\Delta p_0 > 0$ and so $e^{-\beta_S\omega} - e^{-\beta_M|\Gamma|} > 0$ which implies that $-\omega/T_S > - |\Gamma|/T_M$. Similarly, if the query heats the probe then $\Delta p_0 < 0$ and $e^{-\beta_S \omega} - e^{-\beta_M |\Gamma|} < 0$ which implies that $-\omega/T_S < |\Gamma|/T_M$. To summarise we can state the conditions on the query as
\begin{gather}
    \text{Cooling : } \frac{\omega}{T_S} < \frac{|\Gamma|}{T_M} \hspace{1cm}     \text{Heating : } \frac{\omega}{T_S} > \frac{|\Gamma|}{T_M}.
\end{gather}
For the sensitivity condition we require $|\Delta p_0| > c$ with $0 < c < 1-p_0$ giving the constraint
\begin{gather}
 |e^{-\beta_S \omega} - e^{-\beta_M |\Gamma|}| > c \mathcal{Z}_S \mathcal{Z}_f, \label{eq:case}
\end{gather}
which gives two cases, depending on whether the query heats or cools the probe. If the query cools the probe then we have that $e^{-\beta_S\omega} - e^{-\beta_M |\Gamma|} > 0$ so that eq.\eqref{eq:case} leads to  $e^{-\beta_S \omega} - e^{-\beta_M |\Gamma|} > c \mathcal{Z}_S \mathcal{Z}_f$ and rearranging and taking logarithms $
 -\frac{\omega}{T_S} > \log\left( c \mathcal{Z}_S \mathcal{Z}_f  + e^{-\beta_M |\Gamma|}\right)$ so finally
\begin{align}
\text{Sensitive Cooling : }\frac{\omega}{T_S} < -\log\left( c \mathcal{Z}_S \mathcal{Z}_f  + e^{-\beta_M |\Gamma|}\right),
\end{align}
ensures the desired sensitivity if the probe is cooled. Note that this quantity is positive if $ 0< c \mathcal{Z}_S \mathcal{Z}_f  + e^{-\beta_M |\Gamma|} \leq 1$. In the case of heating we have $e^{-\beta_S \omega} - e^{-\beta_M|\Gamma|} < 0$ so that eq.\eqref{eq:case} takes the form $e^{-\beta_M |\Gamma|} - e^{-\beta_S \omega} > c \mathcal{Z}_S \mathcal{Z}_f$ and rearranging and take logarithms leads to $\log\left(e^{-\beta_M |\Gamma|} -  c \mathcal{Z}_S \mathcal{Z}_f\right) >  - \omega/T_S$ and finally
\begin{gather}
\text{Sensitive Heating : }    \frac{\omega}{T_S} > - \log\left(e^{-\beta_M |\Gamma|} -  c \mathcal{Z}_S \mathcal{Z}_f\right),
\end{gather}
which is positive for $c\mathcal{Z}_S\mathcal{Z}_f < e^{-\beta_M |\Gamma|} \leq 1 +c\mathcal{Z}_S\mathcal{Z}_f$. These conditions we have derived for the probe to satisfy the sensitivity we desire for a heat exchange $|\Delta p_0| > c$ can be thought of as tighter versions of the heating and cooling conditions.
\\
\\
Lastly, to make sure we have a well defined temperature after heat exchange we require that the denominator and numerator of the fraction in eq.\eqref{eq:dj_temp} are both positive in the logarithm. Consider
\begin{align}
    \beta'_S = \frac{1}{\omega} \log\left(\frac{1 + \mathcal{Z}^{-1}_f (e^{-\beta_S \omega} - e^{-\beta_M |\Gamma|})}{e^{-\beta_S\omega} - \mathcal{Z}^{-1}_f(e^{-\beta_S\omega} - e^{-\beta_M |\Gamma|})}\right) = \frac{1}{\omega} \log\left(\frac{1 + \mathcal{Z}_S \Delta p_0}{e^{-\beta_S\omega} - \mathcal{Z}_S\Delta p_0}\right), 
\end{align}
in the case of cooling we have that $\Delta p_0 > 0$ so that the numerator is always positive but for the denominator to be positive we require $e^{-\beta_S \omega} > \mathcal{Z}_S\Delta p_0,$ that is $e^{-\beta_S \omega} > \mathcal{Z}^{-1}_f (e^{-\beta_S \omega} - e^{-\beta_M |\Gamma|})$. This leads to 
\begin{align}
-\frac{\omega}{T_S} &> \log\left(\frac{e^{-\beta_S \omega} - e^{-\beta_M |\Gamma|}}{\mathcal{Z}_f}\right)\\
\frac{\omega}{T_S} &< \log\left(\mathcal{Z}_f\right) - \log\left(e^{-\beta_S \omega} - e^{-\beta_M |\Gamma|}\right) 
\end{align}
so that up to $\mathcal{O}\left(\log(e^{-\beta_S \omega} - e^{-\beta_M |\Gamma|})\right)$ for the temperature to be well-defined after a cooling interaction one requires $\omega/T_S < \log(\mathcal{Z}_f)$ where the r.h.s is the free energy of the thermal machine oracle. In the case of heating i.e. $\Delta p_0 < 0$ it is the numerator which can become negative so we require that $1 > \mathcal{Z}_S \Delta p_0$. This leads to $1 > \mathcal{Z}^{-1}_f (e^{-\beta_S \omega} - e^{-\beta_M |\Gamma|})$ and so the condition
\begin{gather}
    \frac{\omega}{T_S} > - \log\left(\mathcal{Z}_f+e^{-\beta_M |\Gamma|}\right)
\end{gather}
which is always true for $\omega/T_S > 0$.
\subsection{Comparing with Classical Probabilistic Approaches to the Deutsch-Jozsa Problem}
\begin{figure}[h]
    \centering
    \includegraphics[width=0.65\linewidth]{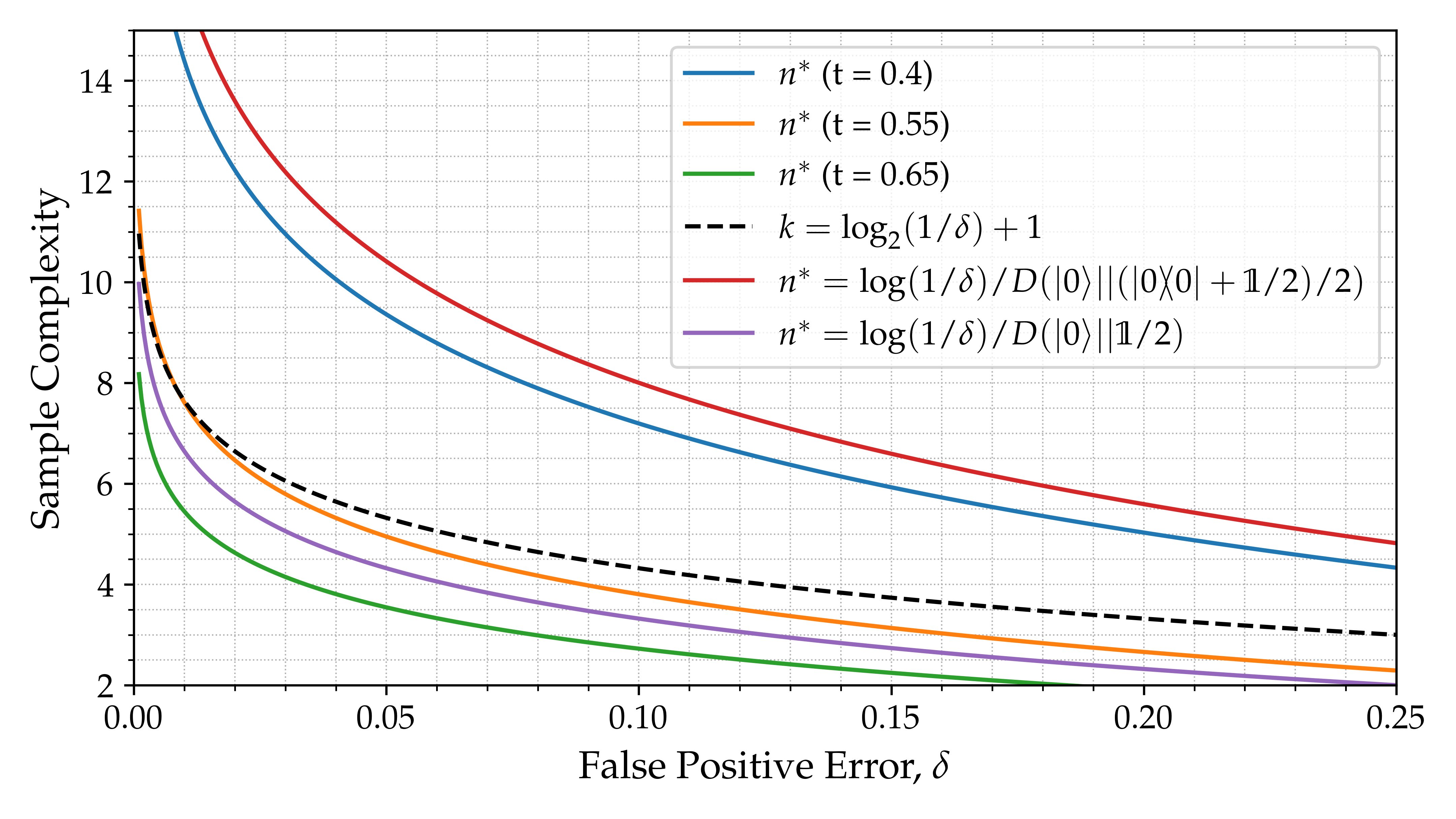}
    \caption{A plot comparing the sample complexity $k$ for probabilistic classical approaches to the Deutsch-Jozsa problem and $n^*$ the sample complexity for different $t$ values, against different false positive error values $\delta$.}
    \label{fig:samp_comp}
\end{figure}
An exponential query complexity separation exists for \textit{deterministic} solutions of the Deutsch-Jozsa problem between classical and quantum approaches. Where, a quantum algorithm can solve the problem in a single query and a classical algorithm requires $2^{n-1} + 1$ queries to determine if an $n$ bit function is balanced or constant. Similarly, we have shown that there is an exponential separation between solving this problem via a quantum thermodynamic heat exchange and the classical solution, but \textit{how does the sample complexity of the presented solution compare with a classical probabilistic approach}?
\\ \\
Let us assume a classical agent is able to randomly sample from a uniform distribution of $n$ bit strings $\{0,1\}^{\times n}$ and use this random sampling to determine whether the Boolean function $f$ is balanced or constant. A strategy the agent might take is as follows, they make a hypothesis that $f(x)$ is constant and wishes to verify this within an error of $\delta$ having sampled $k$ random bit strings  $x \in \{0,1\}^{\times n}$ and examined each of their outputs $f(x)$. That is, $\delta$ is a false positive error where the agent has sampled $k < 2^{n-1}+1$ and not yet found an output which alerts them the function is balanced and so incorrectly thinks the function is balanced.
\\
\\
To obtain an expression for $\delta$ assume that $f$ is balance but that the output of $k$ samples has been constant. With replacement, each sample $x$ has a probability of $1/2$ of $f(x)$ being either 0 or 1 for a balanced $f$. Thus, the probability of getting an output of 0 or 1 $k$ times is $2^{k}$ respectively.  Therefore the false positive probability after $k$ samples is 
\begin{align}
\delta &= P(\text{$k$ 0 outputs}) + P(\text{$k$ 1 outputs})\\
&= 2^{-k} + 2^{-k}\\
&= \frac{1}{2^{k-1}.}
\end{align}
Without replacement i.e. if a bit string $x$ is sampled it is then discarded from the set of $n$ bit strings, we instead have that there are $\binom{2^n}{k}$ ways to select $k$ input bit strings. Since $f$ is balanced it will take half the domain to 0 and half the domain to 1 so that there are $\binom{2^{n-1}}{k}$ ways of obtain each constant outcome respectively. This gives
\begin{align}
\delta' &= P(\text{$k$ 0 outputs}) + P(\text{$k$ 1 outputs})\\
&= \frac{\binom{2^{n-1}}{k} + \binom{2^{n-1}}{k}}{\binom{2^{n}}{k}} = \frac{2\binom{2^{n-1}}{k}}{\binom{2^{n}}{k}}
\end{align}
where $\lim_{k \rightarrow \infty}\delta' = \delta$, that is the false positive error with and without replacement are asymptotically equivalent. Thus asymptotically we see that with this random sampling method we require 
\begin{gather}
    k = \Theta\left(\log_2(1/\delta) + 1\right).
\end{gather}
Having obtained this expression we are now in a position to compare with the asymptotic sample complexity for the heat exchange approach presented in the main text where we had $n^* = \Theta\left(\log(1/\delta)/2t^2\right)$ where $t$ is the distinguishability threshold. In Fig. 6 we give a comparison of the sample complexities for different values of $t$ which suggests that a heat exchange approach could obtain a better sample complexity than the classical probabilistic approach if $t \gtrsim 0.55$. A numerical investigation exploring Fig. 5 for different $t$ values suggests that it could be energetically challenging to obtain $t$ values larger than 0.4 with a qubit probe but it remains on open problem to show a limit on $t$ for probes of arbitrary dimensions and arbitrary unitary heat exchanges beyond the proof of principle model presented in the main text.
\\
\\
We also plot the sample complexity for the probabilistic bipartite heat exchange presented early in the manuscript with sample complexity $n^* = \Theta(\log(1/\delta)/D(\tau_1||(\tau_1 + \tau_2)/2))$. In our model the oracle can at best prepare $\tau_1 = \ket{0}$ and $\tau_2 = \mathbb{1}/2$ leading to red line. But if we allowed the oracle to prepare $\tau_1 = \ket{0}$ and $\tau_2 = \ket{1}$, even this probabilistic bipartite heat exchange would be classical random sampling.
\\
\\
It is important to note that whilst the sample complexity in the classical probabilistic setting stems from sampling the function multiple times for different random inputs, the sample complexity in our model is due to the need to estimate the temperature of the qubit encoding the solution. That is, the solution is already encoded in the state of the probe after a single heat exchange, but reading this solution out requires a constant number of samples/copies of the probe.

\begin{thebibliography}{41}%
	\makeatletter
	\providecommand \@ifxundefined [1]{%
		\@ifx{#1\undefined}
	}%
	\providecommand \@ifnum [1]{%
		\ifnum #1\expandafter \@firstoftwo
		\else \expandafter \@secondoftwo
		\fi
	}%
	\providecommand \@ifx [1]{%
		\ifx #1\expandafter \@firstoftwo
		\else \expandafter \@secondoftwo
		\fi
	}%
	\providecommand \natexlab [1]{#1}%
	\providecommand \enquote  [1]{``#1''}%
	\providecommand \bibnamefont  [1]{#1}%
	\providecommand \bibfnamefont [1]{#1}%
	\providecommand \citenamefont [1]{#1}%
	\providecommand \href@noop [0]{\@secondoftwo}%
	\providecommand \href [0]{\begingroup \@sanitize@url \@href}%
	\providecommand \@href[1]{\@@startlink{#1}\@@href}%
	\providecommand \@@href[1]{\endgroup#1\@@endlink}%
	\providecommand \@sanitize@url [0]{\catcode `\\12\catcode `\$12\catcode `\&12\catcode `\#12\catcode `\^12\catcode `\_12\catcode `\%12\relax}%
	\providecommand \@@startlink[1]{}%
	\providecommand \@@endlink[0]{}%
	\providecommand \url  [0]{\begingroup\@sanitize@url \@url }%
	\providecommand \@url [1]{\endgroup\@href {#1}{\urlprefix }}%
	\providecommand \urlprefix  [0]{URL }%
	\providecommand \Eprint [0]{\href }%
	\providecommand \doibase [0]{https://doi.org/}%
	\providecommand \selectlanguage [0]{\@gobble}%
	\providecommand \bibinfo  [0]{\@secondoftwo}%
	\providecommand \bibfield  [0]{\@secondoftwo}%
	\providecommand \translation [1]{[#1]}%
	\providecommand \BibitemOpen [0]{}%
	\providecommand \bibitemStop [0]{}%
	\providecommand \bibitemNoStop [0]{.\EOS\space}%
	\providecommand \EOS [0]{\spacefactor3000\relax}%
	\providecommand \BibitemShut  [1]{\csname bibitem#1\endcsname}%
	\let\auto@bib@innerbib\@empty
	\bibitem [{\citenamefont {Deutsch}\ and\ \citenamefont {Jozsa}(1992)}]{deutsch1992rapid}%
	\BibitemOpen
	\bibfield  {author} {\bibinfo {author} {\bibfnamefont {D.}~\bibnamefont {Deutsch}}\ and\ \bibinfo {author} {\bibfnamefont {R.}~\bibnamefont {Jozsa}},\ }\bibfield  {title} {\bibinfo {title} {Rapid solution of problems by quantum computation},\ }\href {https://doi.org/10.1098/rspa.1992.0167} {\bibfield  {journal} {\bibinfo  {journal} {Proceedings of the Royal Society of London. Series A: Mathematical and Physical Sciences}\ }\textbf {\bibinfo {volume} {439}},\ \bibinfo {pages} {553} (\bibinfo {year} {1992})}\BibitemShut {NoStop}%
	\bibitem [{\citenamefont {Berthiaume}\ and\ \citenamefont {Brassard}(1992)}]{berthiaume_brassard_92}%
	\BibitemOpen
	\bibfield  {author} {\bibinfo {author} {\bibfnamefont {A.}~\bibnamefont {Berthiaume}}\ and\ \bibinfo {author} {\bibfnamefont {G.}~\bibnamefont {Brassard}},\ }\bibfield  {title} {\bibinfo {title} {The quantum challenge to structural complexity theory},\ }in\ \href {https://doi.org/10.1109/SCT.1992.215388} {\emph {\bibinfo {booktitle} {[1992] Proceedings of the Seventh Annual Structure in Complexity Theory Conference}}}\ (\bibinfo {year} {1992})\ pp.\ \bibinfo {pages} {132--137}\BibitemShut {NoStop}%
	\bibitem [{\citenamefont {Berthiaume}\ and\ \citenamefont {and}(1994)}]{berthiaume_brassard_94}%
	\BibitemOpen
	\bibfield  {author} {\bibinfo {author} {\bibfnamefont {A.}~\bibnamefont {Berthiaume}}\ and\ \bibinfo {author} {\bibfnamefont {G.~B.}\ \bibnamefont {and}},\ }\bibfield  {title} {\bibinfo {title} {Oracle quantum computing},\ }\href {https://doi.org/10.1080/09500349414552351} {\bibfield  {journal} {\bibinfo  {journal} {Journal of Modern Optics}\ }\textbf {\bibinfo {volume} {41}},\ \bibinfo {pages} {2521} (\bibinfo {year} {1994})},\ \Eprint {https://arxiv.org/abs/https://doi.org/10.1080/09500349414552351} {https://doi.org/10.1080/09500349414552351} \BibitemShut {NoStop}%
	\bibitem [{\citenamefont {Bernstein}\ and\ \citenamefont {Vazirani}(1997)}]{bernstein_vazirani}%
	\BibitemOpen
	\bibfield  {author} {\bibinfo {author} {\bibfnamefont {E.}~\bibnamefont {Bernstein}}\ and\ \bibinfo {author} {\bibfnamefont {U.}~\bibnamefont {Vazirani}},\ }\bibfield  {title} {\bibinfo {title} {Quantum complexity theory},\ }\href {https://doi.org/10.1137/S0097539796300921} {\bibfield  {journal} {\bibinfo  {journal} {SIAM Journal on Computing}\ }\textbf {\bibinfo {volume} {26}},\ \bibinfo {pages} {1411} (\bibinfo {year} {1997})},\ \Eprint {https://arxiv.org/abs/https://doi.org/10.1137/S0097539796300921} {https://doi.org/10.1137/S0097539796300921} \BibitemShut {NoStop}%
	\bibitem [{\citenamefont {Watrous}(2009)}]{Watrous2009}%
	\BibitemOpen
	\bibfield  {author} {\bibinfo {author} {\bibfnamefont {J.}~\bibnamefont {Watrous}},\ }\bibinfo {title} {Quantum computational complexity},\ in\ \href {https://doi.org/10.1007/978-0-387-30440-3_428} {\emph {\bibinfo {booktitle} {Encyclopedia of Complexity and Systems Science}}},\ \bibinfo {editor} {edited by\ \bibinfo {editor} {\bibfnamefont {R.~A.}\ \bibnamefont {Meyers}}}\ (\bibinfo  {publisher} {Springer New York},\ \bibinfo {address} {New York, NY},\ \bibinfo {year} {2009})\ pp.\ \bibinfo {pages} {7174--7201}\BibitemShut {NoStop}%
	\bibitem [{\citenamefont {Vinjanampathy}\ and\ \citenamefont {Anders}(2016)}]{anders_review}%
	\BibitemOpen
	\bibfield  {author} {\bibinfo {author} {\bibfnamefont {S.}~\bibnamefont {Vinjanampathy}}\ and\ \bibinfo {author} {\bibfnamefont {J.}~\bibnamefont {Anders}},\ }\bibfield  {title} {\bibinfo {title} {Quantum thermodynamics},\ }\href {https://doi.org/10.1080/00107514.2016.1201896} {\bibfield  {journal} {\bibinfo  {journal} {Contemporary Physics}\ }\textbf {\bibinfo {volume} {57}},\ \bibinfo {pages} {545–579} (\bibinfo {year} {2016})}\BibitemShut {NoStop}%
	\bibitem [{\citenamefont {Goold}\ \emph {et~al.}(2016)\citenamefont {Goold}, \citenamefont {Huber}, \citenamefont {Riera}, \citenamefont {del Rio},\ and\ \citenamefont {Skrzypczyk}}]{Goold_2016}%
	\BibitemOpen
	\bibfield  {author} {\bibinfo {author} {\bibfnamefont {J.}~\bibnamefont {Goold}}, \bibinfo {author} {\bibfnamefont {M.}~\bibnamefont {Huber}}, \bibinfo {author} {\bibfnamefont {A.}~\bibnamefont {Riera}}, \bibinfo {author} {\bibfnamefont {L.}~\bibnamefont {del Rio}},\ and\ \bibinfo {author} {\bibfnamefont {P.}~\bibnamefont {Skrzypczyk}},\ }\bibfield  {title} {\bibinfo {title} {The role of quantum information in thermodynamics—a topical review},\ }\href {https://doi.org/10.1088/1751-8113/49/14/143001} {\bibfield  {journal} {\bibinfo  {journal} {Journal of Physics A: Mathematical and Theoretical}\ }\textbf {\bibinfo {volume} {49}},\ \bibinfo {pages} {143001} (\bibinfo {year} {2016})}\BibitemShut {NoStop}%
	\bibitem [{\citenamefont {Binder}\ \emph {et~al.}(2018)\citenamefont {Binder}, \citenamefont {Correa}, \citenamefont {Gogolin}, \citenamefont {Anders},\ and\ \citenamefont {Adesso}}]{binder2018thermodynamics}%
	\BibitemOpen
	\bibinfo {editor} {\bibfnamefont {F.}~\bibnamefont {Binder}}, \bibinfo {editor} {\bibfnamefont {L.~A.}\ \bibnamefont {Correa}}, \bibinfo {editor} {\bibfnamefont {C.}~\bibnamefont {Gogolin}}, \bibinfo {editor} {\bibfnamefont {J.}~\bibnamefont {Anders}},\ and\ \bibinfo {editor} {\bibfnamefont {G.}~\bibnamefont {Adesso}},\ eds.,\ \href {https://doi.org/10.1007/978-3-319-99046-0} {\emph {\bibinfo {title} {Thermodynamics in the Quantum Regime: Fundamental Aspects and New Directions}}},\ \bibinfo {series} {Fundamental Theories of Physics}, Vol.\ \bibinfo {volume} {195}\ (\bibinfo  {publisher} {Springer International Publishing},\ \bibinfo {address} {Cham},\ \bibinfo {year} {2018})\BibitemShut {NoStop}%
	\bibitem [{\citenamefont {Taranto}\ \emph {et~al.}(2023)\citenamefont {Taranto}, \citenamefont {Bakhshinezhad}, \citenamefont {Bluhm}, \citenamefont {Silva}, \citenamefont {Friis}, \citenamefont {Lock}, \citenamefont {Vitagliano}, \citenamefont {Binder}, \citenamefont {Debarba}, \citenamefont {Schwarzhans}, \citenamefont {Clivaz},\ and\ \citenamefont {Huber}}]{taranto_23}%
	\BibitemOpen
	\bibfield  {author} {\bibinfo {author} {\bibfnamefont {P.}~\bibnamefont {Taranto}}, \bibinfo {author} {\bibfnamefont {F.}~\bibnamefont {Bakhshinezhad}}, \bibinfo {author} {\bibfnamefont {A.}~\bibnamefont {Bluhm}}, \bibinfo {author} {\bibfnamefont {R.}~\bibnamefont {Silva}}, \bibinfo {author} {\bibfnamefont {N.}~\bibnamefont {Friis}}, \bibinfo {author} {\bibfnamefont {M.~P.}\ \bibnamefont {Lock}}, \bibinfo {author} {\bibfnamefont {G.}~\bibnamefont {Vitagliano}}, \bibinfo {author} {\bibfnamefont {F.~C.}\ \bibnamefont {Binder}}, \bibinfo {author} {\bibfnamefont {T.}~\bibnamefont {Debarba}}, \bibinfo {author} {\bibfnamefont {E.}~\bibnamefont {Schwarzhans}}, \bibinfo {author} {\bibfnamefont {F.}~\bibnamefont {Clivaz}},\ and\ \bibinfo {author} {\bibfnamefont {M.}~\bibnamefont {Huber}},\ }\bibfield  {title} {\bibinfo {title} {Landauer versus nernst: What is the true cost of cooling a quantum system?},\ }\href {https://doi.org/10.1103/PRXQuantum.4.010332} {\bibfield  {journal} {\bibinfo  {journal} {PRX Quantum}\
		}\textbf {\bibinfo {volume} {4}},\ \bibinfo {pages} {010332} (\bibinfo {year} {2023})}\BibitemShut {NoStop}%
	\bibitem [{\citenamefont {Hovhannisyan}\ \emph {et~al.}(2013)\citenamefont {Hovhannisyan}, \citenamefont {Perarnau-Llobet}, \citenamefont {Huber},\ and\ \citenamefont {Ac\'{\i}n}}]{hovhanisyan_13}%
	\BibitemOpen
	\bibfield  {author} {\bibinfo {author} {\bibfnamefont {K.~V.}\ \bibnamefont {Hovhannisyan}}, \bibinfo {author} {\bibfnamefont {M.}~\bibnamefont {Perarnau-Llobet}}, \bibinfo {author} {\bibfnamefont {M.}~\bibnamefont {Huber}},\ and\ \bibinfo {author} {\bibfnamefont {A.}~\bibnamefont {Ac\'{\i}n}},\ }\bibfield  {title} {\bibinfo {title} {Entanglement generation is not necessary for optimal work extraction},\ }\href {https://doi.org/10.1103/PhysRevLett.111.240401} {\bibfield  {journal} {\bibinfo  {journal} {Phys. Rev. Lett.}\ }\textbf {\bibinfo {volume} {111}},\ \bibinfo {pages} {240401} (\bibinfo {year} {2013})}\BibitemShut {NoStop}%
	\bibitem [{\citenamefont {Knill}\ and\ \citenamefont {Laflamme}(1998)}]{DQC1}%
	\BibitemOpen
	\bibfield  {author} {\bibinfo {author} {\bibfnamefont {E.}~\bibnamefont {Knill}}\ and\ \bibinfo {author} {\bibfnamefont {R.}~\bibnamefont {Laflamme}},\ }\bibfield  {title} {\bibinfo {title} {Power of one bit of quantum information},\ }\href {https://doi.org/10.1103/physrevlett.81.5672} {\bibfield  {journal} {\bibinfo  {journal} {Physical Review Letters}\ }\textbf {\bibinfo {volume} {81}},\ \bibinfo {pages} {5672} (\bibinfo {year} {1998})}\BibitemShut {NoStop}%
	\bibitem [{\citenamefont {Ambainis}\ \emph {et~al.}(2006)\citenamefont {Ambainis}, \citenamefont {Schulman},\ and\ \citenamefont {Vazirani}}]{ambainis2000computing}%
	\BibitemOpen
	\bibfield  {author} {\bibinfo {author} {\bibfnamefont {A.}~\bibnamefont {Ambainis}}, \bibinfo {author} {\bibfnamefont {L.~J.}\ \bibnamefont {Schulman}},\ and\ \bibinfo {author} {\bibfnamefont {U.}~\bibnamefont {Vazirani}},\ }\bibfield  {title} {\bibinfo {title} {Computing with highly mixed states},\ }\href {https://doi.org/10.1145/1147954.1147962} {\bibfield  {journal} {\bibinfo  {journal} {Journal of the ACM}\ }\textbf {\bibinfo {volume} {53}},\ \bibinfo {pages} {507–531} (\bibinfo {year} {2006})}\BibitemShut {NoStop}%
	\bibitem [{\citenamefont {Laflamme}\ \emph {et~al.}(2002)\citenamefont {Laflamme}, \citenamefont {Knill}, \citenamefont {Cory}, \citenamefont {Fortunato}, \citenamefont {Havel}, \citenamefont {Miquel}, \citenamefont {Martinez}, \citenamefont {Negrevergne}, \citenamefont {Ortiz}, \citenamefont {Pravia} \emph {et~al.}}]{laflamme2002introduction}%
	\BibitemOpen
	\bibfield  {author} {\bibinfo {author} {\bibfnamefont {R.}~\bibnamefont {Laflamme}}, \bibinfo {author} {\bibfnamefont {E.}~\bibnamefont {Knill}}, \bibinfo {author} {\bibfnamefont {D.}~\bibnamefont {Cory}}, \bibinfo {author} {\bibfnamefont {E.}~\bibnamefont {Fortunato}}, \bibinfo {author} {\bibfnamefont {T.}~\bibnamefont {Havel}}, \bibinfo {author} {\bibfnamefont {C.}~\bibnamefont {Miquel}}, \bibinfo {author} {\bibfnamefont {R.}~\bibnamefont {Martinez}}, \bibinfo {author} {\bibfnamefont {C.}~\bibnamefont {Negrevergne}}, \bibinfo {author} {\bibfnamefont {G.}~\bibnamefont {Ortiz}}, \bibinfo {author} {\bibfnamefont {M.}~\bibnamefont {Pravia}}, \emph {et~al.},\ }\bibfield  {title} {\bibinfo {title} {Introduction to nmr quantum information processing},\ }\href@noop {} {\bibfield  {journal} {\bibinfo  {journal} {arXiv preprint quant-ph/0207172}\ } (\bibinfo {year} {2002})}\BibitemShut {NoStop}%
	\bibitem [{\citenamefont {Morimae}\ \emph {et~al.}(2017)\citenamefont {Morimae}, \citenamefont {Fujii},\ and\ \citenamefont {Nishimura}}]{non_clean}%
	\BibitemOpen
	\bibfield  {author} {\bibinfo {author} {\bibfnamefont {T.}~\bibnamefont {Morimae}}, \bibinfo {author} {\bibfnamefont {K.}~\bibnamefont {Fujii}},\ and\ \bibinfo {author} {\bibfnamefont {H.}~\bibnamefont {Nishimura}},\ }\bibfield  {title} {\bibinfo {title} {Power of one nonclean qubit},\ }\bibfield  {journal} {\bibinfo  {journal} {Physical Review A}\ }\textbf {\bibinfo {volume} {95}},\ \href {https://doi.org/10.1103/physreva.95.042336} {10.1103/physreva.95.042336} (\bibinfo {year} {2017})\BibitemShut {NoStop}%
	\bibitem [{\citenamefont {Poulin}\ \emph {et~al.}(2003)\citenamefont {Poulin}, \citenamefont {Laflamme}, \citenamefont {Milburn},\ and\ \citenamefont {Paz}}]{integrability}%
	\BibitemOpen
	\bibfield  {author} {\bibinfo {author} {\bibfnamefont {D.}~\bibnamefont {Poulin}}, \bibinfo {author} {\bibfnamefont {R.}~\bibnamefont {Laflamme}}, \bibinfo {author} {\bibfnamefont {G.~J.}\ \bibnamefont {Milburn}},\ and\ \bibinfo {author} {\bibfnamefont {J.~P.}\ \bibnamefont {Paz}},\ }\bibfield  {title} {\bibinfo {title} {Testing integrability with a single bit of quantum information},\ }\href {https://doi.org/10.1103/PhysRevA.68.022302} {\bibfield  {journal} {\bibinfo  {journal} {Phys. Rev. A}\ }\textbf {\bibinfo {volume} {68}},\ \bibinfo {pages} {022302} (\bibinfo {year} {2003})}\BibitemShut {NoStop}%
	\bibitem [{\citenamefont {Poulin}\ \emph {et~al.}(2004)\citenamefont {Poulin}, \citenamefont {Blume-Kohout}, \citenamefont {Laflamme},\ and\ \citenamefont {Ollivier}}]{poulin_2}%
	\BibitemOpen
	\bibfield  {author} {\bibinfo {author} {\bibfnamefont {D.}~\bibnamefont {Poulin}}, \bibinfo {author} {\bibfnamefont {R.}~\bibnamefont {Blume-Kohout}}, \bibinfo {author} {\bibfnamefont {R.}~\bibnamefont {Laflamme}},\ and\ \bibinfo {author} {\bibfnamefont {H.}~\bibnamefont {Ollivier}},\ }\bibfield  {title} {\bibinfo {title} {Exponential speedup with a single bit of quantum information: Measuring the average fidelity decay},\ }\href {https://doi.org/10.1103/PhysRevLett.92.177906} {\bibfield  {journal} {\bibinfo  {journal} {Phys. Rev. Lett.}\ }\textbf {\bibinfo {volume} {92}},\ \bibinfo {pages} {177906} (\bibinfo {year} {2004})}\BibitemShut {NoStop}%
	\bibitem [{\citenamefont {Shepherd}(2006)}]{shepherd2006computation}%
	\BibitemOpen
	\bibfield  {author} {\bibinfo {author} {\bibfnamefont {D.}~\bibnamefont {Shepherd}},\ }\bibfield  {title} {\bibinfo {title} {Computation with unitaries and one pure qubit},\ }\href@noop {} {\bibfield  {journal} {\bibinfo  {journal} {arXiv preprint quant-ph/0608132}\ } (\bibinfo {year} {2006})}\BibitemShut {NoStop}%
	\bibitem [{\citenamefont {Shor}\ and\ \citenamefont {Jordan}(2008)}]{shor2008estimating}%
	\BibitemOpen
	\bibfield  {author} {\bibinfo {author} {\bibfnamefont {P.}~\bibnamefont {Shor}}\ and\ \bibinfo {author} {\bibfnamefont {S.}~\bibnamefont {Jordan}},\ }\bibfield  {title} {\bibinfo {title} {Estimating jones polynomials is a complete problem for one clean qubit},\ }\href {https://doi.org/10.26421/qic8.8-9-1} {\bibfield  {journal} {\bibinfo  {journal} {Quantum Information and Computation}\ }\textbf {\bibinfo {volume} {8}},\ \bibinfo {pages} {681–714} (\bibinfo {year} {2008})}\BibitemShut {NoStop}%
	\bibitem [{\citenamefont {Aifer}\ \emph {et~al.}(2024)\citenamefont {Aifer}, \citenamefont {Donatella}, \citenamefont {Gordon}, \citenamefont {Duffield}, \citenamefont {Ahle}, \citenamefont {Simpson}, \citenamefont {Crooks},\ and\ \citenamefont {Coles}}]{aifer2024thermodynamic}%
	\BibitemOpen
	\bibfield  {author} {\bibinfo {author} {\bibfnamefont {M.}~\bibnamefont {Aifer}}, \bibinfo {author} {\bibfnamefont {K.}~\bibnamefont {Donatella}}, \bibinfo {author} {\bibfnamefont {M.~H.}\ \bibnamefont {Gordon}}, \bibinfo {author} {\bibfnamefont {S.}~\bibnamefont {Duffield}}, \bibinfo {author} {\bibfnamefont {T.}~\bibnamefont {Ahle}}, \bibinfo {author} {\bibfnamefont {D.}~\bibnamefont {Simpson}}, \bibinfo {author} {\bibfnamefont {G.~E.}\ \bibnamefont {Crooks}},\ and\ \bibinfo {author} {\bibfnamefont {P.~J.}\ \bibnamefont {Coles}},\ }\bibfield  {title} {\bibinfo {title} {Thermodynamic linear algebra},\ }\href {https://doi.org/10.1038/s44335-024-00014-0} {\bibfield  {journal} {\bibinfo  {journal} {npj Unconventional Computing}\ }\textbf {\bibinfo {volume} {1}},\ \bibinfo {pages} {13} (\bibinfo {year} {2024})}\BibitemShut {NoStop}%
	\bibitem [{\citenamefont {Bartosik}\ \emph {et~al.}(2024)\citenamefont {Bartosik}, \citenamefont {Donatella}, \citenamefont {Aifer}, \citenamefont {Melanson}, \citenamefont {Perarnau-Llobet}, \citenamefont {Brunner},\ and\ \citenamefont {Coles}}]{bartosik2024thermodynamicalgorithmsquadraticprogramming}%
	\BibitemOpen
	\bibfield  {author} {\bibinfo {author} {\bibfnamefont {P.-L.}\ \bibnamefont {Bartosik}}, \bibinfo {author} {\bibfnamefont {K.}~\bibnamefont {Donatella}}, \bibinfo {author} {\bibfnamefont {M.}~\bibnamefont {Aifer}}, \bibinfo {author} {\bibfnamefont {D.}~\bibnamefont {Melanson}}, \bibinfo {author} {\bibfnamefont {M.}~\bibnamefont {Perarnau-Llobet}}, \bibinfo {author} {\bibfnamefont {N.}~\bibnamefont {Brunner}},\ and\ \bibinfo {author} {\bibfnamefont {P.~J.}\ \bibnamefont {Coles}},\ }\href {https://arxiv.org/abs/2411.14224} {\bibinfo {title} {Thermodynamic algorithms for quadratic programming}} (\bibinfo {year} {2024}),\ \Eprint {https://arxiv.org/abs/2411.14224} {arXiv:2411.14224 [cs.ET]} \BibitemShut {NoStop}%
	\bibitem [{\citenamefont {Linden}\ \emph {et~al.}(2010)\citenamefont {Linden}, \citenamefont {Popescu},\ and\ \citenamefont {Skrzypczyk}}]{skrzypczyk_10}%
	\BibitemOpen
	\bibfield  {author} {\bibinfo {author} {\bibfnamefont {N.}~\bibnamefont {Linden}}, \bibinfo {author} {\bibfnamefont {S.}~\bibnamefont {Popescu}},\ and\ \bibinfo {author} {\bibfnamefont {P.}~\bibnamefont {Skrzypczyk}},\ }\bibfield  {title} {\bibinfo {title} {How small can thermal machines be? the smallest possible refrigerator},\ }\href {https://doi.org/10.1103/PhysRevLett.105.130401} {\bibfield  {journal} {\bibinfo  {journal} {Phys. Rev. Lett.}\ }\textbf {\bibinfo {volume} {105}},\ \bibinfo {pages} {130401} (\bibinfo {year} {2010})}\BibitemShut {NoStop}%
	\bibitem [{\citenamefont {Levy}\ and\ \citenamefont {Kosloff}(2012)}]{ronnie_12}%
	\BibitemOpen
	\bibfield  {author} {\bibinfo {author} {\bibfnamefont {A.}~\bibnamefont {Levy}}\ and\ \bibinfo {author} {\bibfnamefont {R.}~\bibnamefont {Kosloff}},\ }\bibfield  {title} {\bibinfo {title} {Quantum absorption refrigerator},\ }\href {https://doi.org/10.1103/PhysRevLett.108.070604} {\bibfield  {journal} {\bibinfo  {journal} {Phys. Rev. Lett.}\ }\textbf {\bibinfo {volume} {108}},\ \bibinfo {pages} {070604} (\bibinfo {year} {2012})}\BibitemShut {NoStop}%
	\bibitem [{\citenamefont {Silva}\ \emph {et~al.}(2016)\citenamefont {Silva}, \citenamefont {Manzano}, \citenamefont {Skrzypczyk},\ and\ \citenamefont {Brunner}}]{ralph_swap}%
	\BibitemOpen
	\bibfield  {author} {\bibinfo {author} {\bibfnamefont {R.}~\bibnamefont {Silva}}, \bibinfo {author} {\bibfnamefont {G.}~\bibnamefont {Manzano}}, \bibinfo {author} {\bibfnamefont {P.}~\bibnamefont {Skrzypczyk}},\ and\ \bibinfo {author} {\bibfnamefont {N.}~\bibnamefont {Brunner}},\ }\bibfield  {title} {\bibinfo {title} {Performance of autonomous quantum thermal machines: Hilbert space dimension as a thermodynamical resource},\ }\href {https://doi.org/10.1103/PhysRevE.94.032120} {\bibfield  {journal} {\bibinfo  {journal} {Phys. Rev. E}\ }\textbf {\bibinfo {volume} {94}},\ \bibinfo {pages} {032120} (\bibinfo {year} {2016})}\BibitemShut {NoStop}%
	\bibitem [{\citenamefont {Mitchison}(2019)}]{Mitchison_review}%
	\BibitemOpen
	\bibfield  {author} {\bibinfo {author} {\bibfnamefont {M.~T.}\ \bibnamefont {Mitchison}},\ }\bibfield  {title} {\bibinfo {title} {Quantum thermal absorption machines: refrigerators, engines and clocks},\ }\href {https://doi.org/10.1080/00107514.2019.1631555} {\bibfield  {journal} {\bibinfo  {journal} {Contemporary Physics}\ }\textbf {\bibinfo {volume} {60}},\ \bibinfo {pages} {164} (\bibinfo {year} {2019})}\BibitemShut {NoStop}%
	\bibitem [{Note1()}]{Note1}%
	\BibitemOpen
	\bibinfo {note} {We will examine the no. of samples in the \protect \textit {Reading Out} section.}\BibitemShut {Stop}%
	\bibitem [{\citenamefont {Xuereb}(2025{\natexlab{a}})}]{sup_mat}%
	\BibitemOpen
	\bibfield  {author} {\bibinfo {author} {\bibfnamefont {J.}~\bibnamefont {Xuereb}},\ }\href {https://doi.org/To be added by publisher} {\bibinfo {title} {Supplemental material : A temperature change can solve the deutsch-jozsa problem}} (\bibinfo {year} {2025}{\natexlab{a}})\BibitemShut {NoStop}%
	\bibitem [{\citenamefont {Brunner}\ \emph {et~al.}(2012)\citenamefont {Brunner}, \citenamefont {Linden}, \citenamefont {Popescu},\ and\ \citenamefont {Skrzypczyk}}]{brunner_12}%
	\BibitemOpen
	\bibfield  {author} {\bibinfo {author} {\bibfnamefont {N.}~\bibnamefont {Brunner}}, \bibinfo {author} {\bibfnamefont {N.}~\bibnamefont {Linden}}, \bibinfo {author} {\bibfnamefont {S.}~\bibnamefont {Popescu}},\ and\ \bibinfo {author} {\bibfnamefont {P.}~\bibnamefont {Skrzypczyk}},\ }\bibfield  {title} {\bibinfo {title} {Virtual qubits, virtual temperatures, and the foundations of thermodynamics},\ }\href {https://doi.org/10.1103/PhysRevE.85.051117} {\bibfield  {journal} {\bibinfo  {journal} {Phys. Rev. E}\ }\textbf {\bibinfo {volume} {85}},\ \bibinfo {pages} {051117} (\bibinfo {year} {2012})}\BibitemShut {NoStop}%
	\bibitem [{\citenamefont {Clivaz}\ \emph {et~al.}(2019)\citenamefont {Clivaz}, \citenamefont {Silva}, \citenamefont {Haack}, \citenamefont {Brask}, \citenamefont {Brunner},\ and\ \citenamefont {Huber}}]{fab_limits}%
	\BibitemOpen
	\bibfield  {author} {\bibinfo {author} {\bibfnamefont {F.}~\bibnamefont {Clivaz}}, \bibinfo {author} {\bibfnamefont {R.}~\bibnamefont {Silva}}, \bibinfo {author} {\bibfnamefont {G.}~\bibnamefont {Haack}}, \bibinfo {author} {\bibfnamefont {J.~B.}\ \bibnamefont {Brask}}, \bibinfo {author} {\bibfnamefont {N.}~\bibnamefont {Brunner}},\ and\ \bibinfo {author} {\bibfnamefont {M.}~\bibnamefont {Huber}},\ }\bibfield  {title} {\bibinfo {title} {Unifying paradigms of quantum refrigeration: A universal and attainable bound on cooling},\ }\href {https://doi.org/10.1103/PhysRevLett.123.170605} {\bibfield  {journal} {\bibinfo  {journal} {Phys. Rev. Lett.}\ }\textbf {\bibinfo {volume} {123}},\ \bibinfo {pages} {170605} (\bibinfo {year} {2019})}\BibitemShut {NoStop}%
	\bibitem [{\citenamefont {Erker}\ \emph {et~al.}(2017)\citenamefont {Erker}, \citenamefont {Mitchison}, \citenamefont {Silva}, \citenamefont {Woods}, \citenamefont {Brunner},\ and\ \citenamefont {Huber}}]{erker_17}%
	\BibitemOpen
	\bibfield  {author} {\bibinfo {author} {\bibfnamefont {P.}~\bibnamefont {Erker}}, \bibinfo {author} {\bibfnamefont {M.~T.}\ \bibnamefont {Mitchison}}, \bibinfo {author} {\bibfnamefont {R.}~\bibnamefont {Silva}}, \bibinfo {author} {\bibfnamefont {M.~P.}\ \bibnamefont {Woods}}, \bibinfo {author} {\bibfnamefont {N.}~\bibnamefont {Brunner}},\ and\ \bibinfo {author} {\bibfnamefont {M.}~\bibnamefont {Huber}},\ }\bibfield  {title} {\bibinfo {title} {Autonomous quantum clocks: Does thermodynamics limit our ability to measure time?},\ }\href {https://doi.org/10.1103/PhysRevX.7.031022} {\bibfield  {journal} {\bibinfo  {journal} {Phys. Rev. X}\ }\textbf {\bibinfo {volume} {7}},\ \bibinfo {pages} {031022} (\bibinfo {year} {2017})}\BibitemShut {NoStop}%
	\bibitem [{\citenamefont {Hofer}\ \emph {et~al.}(2017)\citenamefont {Hofer}, \citenamefont {Brask}, \citenamefont {Perarnau-Llobet},\ and\ \citenamefont {Brunner}}]{hofer_17}%
	\BibitemOpen
	\bibfield  {author} {\bibinfo {author} {\bibfnamefont {P.~P.}\ \bibnamefont {Hofer}}, \bibinfo {author} {\bibfnamefont {J.~B.}\ \bibnamefont {Brask}}, \bibinfo {author} {\bibfnamefont {M.}~\bibnamefont {Perarnau-Llobet}},\ and\ \bibinfo {author} {\bibfnamefont {N.}~\bibnamefont {Brunner}},\ }\bibfield  {title} {\bibinfo {title} {Quantum thermal machine as a thermometer},\ }\href {https://doi.org/10.1103/PhysRevLett.119.090603} {\bibfield  {journal} {\bibinfo  {journal} {Phys. Rev. Lett.}\ }\textbf {\bibinfo {volume} {119}},\ \bibinfo {pages} {090603} (\bibinfo {year} {2017})}\BibitemShut {NoStop}%
	\bibitem [{\citenamefont {Mehboudi}\ \emph {et~al.}(2019)\citenamefont {Mehboudi}, \citenamefont {Sanpera},\ and\ \citenamefont {Correa}}]{Mehboudi_2019}%
	\BibitemOpen
	\bibfield  {author} {\bibinfo {author} {\bibfnamefont {M.}~\bibnamefont {Mehboudi}}, \bibinfo {author} {\bibfnamefont {A.}~\bibnamefont {Sanpera}},\ and\ \bibinfo {author} {\bibfnamefont {L.~A.}\ \bibnamefont {Correa}},\ }\bibfield  {title} {\bibinfo {title} {Thermometry in the quantum regime: recent theoretical progress},\ }\href {https://doi.org/10.1088/1751-8121/ab2828} {\bibfield  {journal} {\bibinfo  {journal} {Journal of Physics A: Mathematical and Theoretical}\ }\textbf {\bibinfo {volume} {52}},\ \bibinfo {pages} {303001} (\bibinfo {year} {2019})}\BibitemShut {NoStop}%
	\bibitem [{\citenamefont {Thomas M.~Cover}(2005)}]{thomas_cover}%
	\BibitemOpen
	\bibfield  {author} {\bibinfo {author} {\bibfnamefont {J.~A.~T.}\ \bibnamefont {Thomas M.~Cover}},\ }\bibinfo {title} {Information theory and statistics},\ in\ \href {https://doi.org/https://doi.org/10.1002/047174882X.ch11} {\emph {\bibinfo {booktitle} {Elements of Information Theory}}}\ (\bibinfo  {publisher} {John Wiley \& Sons, Ltd},\ \bibinfo {year} {2005})\ Chap.~\bibinfo {chapter} {11}, pp.\ \bibinfo {pages} {347--408}\BibitemShut {NoStop}%
	\bibitem [{\citenamefont {Nielsen}\ and\ \citenamefont {Chuang}(2010)}]{Nielsen_Chuang_2010}%
	\BibitemOpen
	\bibfield  {author} {\bibinfo {author} {\bibfnamefont {M.~A.}\ \bibnamefont {Nielsen}}\ and\ \bibinfo {author} {\bibfnamefont {I.~L.}\ \bibnamefont {Chuang}},\ }\href@noop {} {\emph {\bibinfo {title} {Quantum Computation and Quantum Information: 10th Anniversary Edition}}}\ (\bibinfo  {publisher} {Cambridge University Press},\ \bibinfo {year} {2010})\BibitemShut {NoStop}%
	\bibitem [{\citenamefont {Almanza-Marrero}\ and\ \citenamefont {Manzano}(2025)}]{almanzamarrero2025certifyingquantumenhancementsthermal}%
	\BibitemOpen
	\bibfield  {author} {\bibinfo {author} {\bibfnamefont {J.~A.}\ \bibnamefont {Almanza-Marrero}}\ and\ \bibinfo {author} {\bibfnamefont {G.}~\bibnamefont {Manzano}},\ }\href {https://arxiv.org/abs/2403.19280} {\bibinfo {title} {Certifying quantum enhancements in thermal machines beyond the thermodynamic uncertainty relation}} (\bibinfo {year} {2025}),\ \Eprint {https://arxiv.org/abs/2403.19280} {arXiv:2403.19280 [quant-ph]} \BibitemShut {NoStop}%
	\bibitem [{\citenamefont {Xuereb}(2025{\natexlab{b}})}]{code}%
	\BibitemOpen
	\bibfield  {author} {\bibinfo {author} {\bibfnamefont {J.}~\bibnamefont {Xuereb}},\ }\href {https://github.com/jqed-xuereb/A-Temperature-Change-can-Solve-the-Deutsch---Josza-Problem} {\bibinfo {title} {https://github.com/jqed-xuereb/a-temperature-change-can-solve-the-deutsch---josza-problem}} (\bibinfo {year} {2025}{\natexlab{b}})\BibitemShut {NoStop}%
	\bibitem [{\citenamefont {Pekola}(2015)}]{Pekola2015}%
	\BibitemOpen
	\bibfield  {author} {\bibinfo {author} {\bibfnamefont {J.}~\bibnamefont {Pekola}},\ }\bibfield  {title} {\bibinfo {title} {Towards quantum thermodynamics in electronic circuits},\ }\href {https://doi.org/10.1038/nphys3169} {\bibfield  {journal} {\bibinfo  {journal} {Nature Physics}\ }\textbf {\bibinfo {volume} {11}},\ \bibinfo {pages} {118} (\bibinfo {year} {2015})}\BibitemShut {NoStop}%
	\bibitem [{\citenamefont {van~der Wiel}\ \emph {et~al.}(2002)\citenamefont {van~der Wiel}, \citenamefont {De~Franceschi}, \citenamefont {Elzerman}, \citenamefont {Fujisawa}, \citenamefont {Tarucha},\ and\ \citenamefont {Kouwenhoven}}]{rmp_trans_02}%
	\BibitemOpen
	\bibfield  {author} {\bibinfo {author} {\bibfnamefont {W.~G.}\ \bibnamefont {van~der Wiel}}, \bibinfo {author} {\bibfnamefont {S.}~\bibnamefont {De~Franceschi}}, \bibinfo {author} {\bibfnamefont {J.~M.}\ \bibnamefont {Elzerman}}, \bibinfo {author} {\bibfnamefont {T.}~\bibnamefont {Fujisawa}}, \bibinfo {author} {\bibfnamefont {S.}~\bibnamefont {Tarucha}},\ and\ \bibinfo {author} {\bibfnamefont {L.~P.}\ \bibnamefont {Kouwenhoven}},\ }\bibfield  {title} {\bibinfo {title} {Electron transport through double quantum dots},\ }\href {https://doi.org/10.1103/RevModPhys.75.1} {\bibfield  {journal} {\bibinfo  {journal} {Rev. Mod. Phys.}\ }\textbf {\bibinfo {volume} {75}},\ \bibinfo {pages} {1} (\bibinfo {year} {2002})}\BibitemShut {NoStop}%
	\bibitem [{\citenamefont {Hanson}\ \emph {et~al.}(2007)\citenamefont {Hanson}, \citenamefont {Kouwenhoven}, \citenamefont {Petta}, \citenamefont {Tarucha},\ and\ \citenamefont {Vandersypen}}]{rmp_spin_dot_07}%
	\BibitemOpen
	\bibfield  {author} {\bibinfo {author} {\bibfnamefont {R.}~\bibnamefont {Hanson}}, \bibinfo {author} {\bibfnamefont {L.~P.}\ \bibnamefont {Kouwenhoven}}, \bibinfo {author} {\bibfnamefont {J.~R.}\ \bibnamefont {Petta}}, \bibinfo {author} {\bibfnamefont {S.}~\bibnamefont {Tarucha}},\ and\ \bibinfo {author} {\bibfnamefont {L.~M.~K.}\ \bibnamefont {Vandersypen}},\ }\bibfield  {title} {\bibinfo {title} {Spins in few-electron quantum dots},\ }\href {https://doi.org/10.1103/RevModPhys.79.1217} {\bibfield  {journal} {\bibinfo  {journal} {Rev. Mod. Phys.}\ }\textbf {\bibinfo {volume} {79}},\ \bibinfo {pages} {1217} (\bibinfo {year} {2007})}\BibitemShut {NoStop}%
	\bibitem [{\citenamefont {Zwanenburg}\ \emph {et~al.}(2013)\citenamefont {Zwanenburg}, \citenamefont {Dzurak}, \citenamefont {Morello}, \citenamefont {Simmons}, \citenamefont {Hollenberg}, \citenamefont {Klimeck}, \citenamefont {Rogge}, \citenamefont {Coppersmith},\ and\ \citenamefont {Eriksson}}]{dots_rev_mod}%
	\BibitemOpen
	\bibfield  {author} {\bibinfo {author} {\bibfnamefont {F.~A.}\ \bibnamefont {Zwanenburg}}, \bibinfo {author} {\bibfnamefont {A.~S.}\ \bibnamefont {Dzurak}}, \bibinfo {author} {\bibfnamefont {A.}~\bibnamefont {Morello}}, \bibinfo {author} {\bibfnamefont {M.~Y.}\ \bibnamefont {Simmons}}, \bibinfo {author} {\bibfnamefont {L.~C.~L.}\ \bibnamefont {Hollenberg}}, \bibinfo {author} {\bibfnamefont {G.}~\bibnamefont {Klimeck}}, \bibinfo {author} {\bibfnamefont {S.}~\bibnamefont {Rogge}}, \bibinfo {author} {\bibfnamefont {S.~N.}\ \bibnamefont {Coppersmith}},\ and\ \bibinfo {author} {\bibfnamefont {M.~A.}\ \bibnamefont {Eriksson}},\ }\bibfield  {title} {\bibinfo {title} {Silicon quantum electronics},\ }\href {https://doi.org/10.1103/RevModPhys.85.961} {\bibfield  {journal} {\bibinfo  {journal} {Rev. Mod. Phys.}\ }\textbf {\bibinfo {volume} {85}},\ \bibinfo {pages} {961} (\bibinfo {year} {2013})}\BibitemShut {NoStop}%
	\bibitem [{\citenamefont {Blais}\ \emph {et~al.}(2021)\citenamefont {Blais}, \citenamefont {Grimsmo}, \citenamefont {Girvin},\ and\ \citenamefont {Wallraff}}]{rmp_transmon}%
	\BibitemOpen
	\bibfield  {author} {\bibinfo {author} {\bibfnamefont {A.}~\bibnamefont {Blais}}, \bibinfo {author} {\bibfnamefont {A.~L.}\ \bibnamefont {Grimsmo}}, \bibinfo {author} {\bibfnamefont {S.~M.}\ \bibnamefont {Girvin}},\ and\ \bibinfo {author} {\bibfnamefont {A.}~\bibnamefont {Wallraff}},\ }\bibfield  {title} {\bibinfo {title} {Circuit quantum electrodynamics},\ }\href {https://doi.org/10.1103/RevModPhys.93.025005} {\bibfield  {journal} {\bibinfo  {journal} {Rev. Mod. Phys.}\ }\textbf {\bibinfo {volume} {93}},\ \bibinfo {pages} {025005} (\bibinfo {year} {2021})}\BibitemShut {NoStop}%
	\bibitem [{\citenamefont {Aamir}\ \emph {et~al.}(2025)\citenamefont {Aamir}, \citenamefont {Suria}, \citenamefont {Guzm{\'a}n} \emph {et~al.}}]{Aamir2025}%
	\BibitemOpen
	\bibfield  {author} {\bibinfo {author} {\bibfnamefont {M.~A.}\ \bibnamefont {Aamir}}, \bibinfo {author} {\bibfnamefont {P.~J.}\ \bibnamefont {Suria}}, \bibinfo {author} {\bibfnamefont {J.~A.~M.}\ \bibnamefont {Guzm{\'a}n}}, \emph {et~al.},\ }\bibfield  {title} {\bibinfo {title} {Thermally driven quantum refrigerator autonomously resets a superconducting qubit},\ }\href {https://doi.org/10.1038/s41567-024-02708-5} {\bibfield  {journal} {\bibinfo  {journal} {Nature Physics}\ }\textbf {\bibinfo {volume} {21}},\ \bibinfo {pages} {318} (\bibinfo {year} {2025})}\BibitemShut {NoStop}%
\end{thebibliography}
\end{document}